\documentclass{ws-ijmpb}
\usepackage{epsfig}
\begin{document}
\markboth{M.N.Kiselev}{Semi-fermionic representation for spin
systems ...}
\author{{\large M.N.Kiselev}}
\address{Institut f\"ur Theoretische Physik und Astrophysik, Universit\"at W\"urzburg,
D-97074 W\"urzburg, Germany\\}
\date{\today}

\title{Semi-fermionic representation for spin systems under equilibrium and non-equilibrium conditions}
\maketitle

\begin{history}
\received{\today}
\revised{(revised date)}
\end{history}

\begin{abstract}
We present a general derivation of semi-fermionic representation
for spin operators in terms of a bilinear combination of fermions
in real and imaginary time formalisms. The constraint on fermionic
occupation numbers is fulfilled by means of imaginary Lagrange
multipliers resulting in special shape of quasiparticle
distribution functions. We show how Schwinger-Keldysh technique
for spin operators is constructed with the help of semi-fermions.
We demonstrate how the idea of semi-fermionic representation might
be extended to the groups possessing dynamic symmetries (e.g.
singlet/triplet transitions in quantum dots). We illustrate the
application
of semi-fermionic representations for various problems of strongly correlated and mesoscopic physics.\\
\mbox{}\\
\mbox{}
\end{abstract}

\section*{Introduction}
It is known that spin operators satisfy neither Fermi nor Bose commutation relations. For example,
the Pauli matrices for $S=1/2$ operator commute on different sites and anticommute on the same site.
The commutation relations for spins are determined by $SU(2)$ algebra, leading to the absence of a
Wick theorem for the generators. To avoid this difficulty and construct a diagrammatic
technique and path integral representation for spin systems various approaches have been used.
The first class of approaches is based on representation of spins as bilinear combination of
Fermi or Bose operators \cite{holstein40a}-\cite{larkin68b}, whereas the representations
belonging to the second class deal with
more complex objects like, e.g. the Hubbard \cite{hubbard65a} and supersymmetric
\cite{coleman00a} operators, the
nonlinear sigma model \cite{book2} etc. However, in all cases
the fundamental problem which is  at the heart of the difficulty is the local constraint problem.
To illustrate it, let's consider e.g., first class of representations. Introducing the
auxiliary Fermi or Bose fields
makes the dimensionality of the Hilbert space, where these operators act, greater than the
dimensionality of the Hilbert space for the spin operators. As a result, the spurious unphysical states
should be excluded from the consideration which leads in turn  to some restrictions (constraints)
on bilinear combinations of Fermi/Bose operators, resulting in substantial complication
of corresponding rules of the diagrammatic technique.
The representations from the second class suffer from the
same kind of problem, transformed either into a high nonlinearity of resulting model
(non-linear sigma model) or hierarchical structure of perturbation series in the absence of Wick theorem
(Hubbard operators). The exclusion of double occupied and empty states for a $S=1/2$ impurity
interacting with conduction electron bath (single impurity Kondo model), is controlled by fictitious
chemical potential (Lagrange multiplier) of Abrikosov pseudofermions \cite{abrikosov65a}.
At the end of calculations
this ``chemical potential'' $\lambda$ should be put $\lambda \to -\infty$ to ``freeze out''
all unphysical states. In other words, there exists an additional $U(1)$ gauge field which freezes
the charge fluctuations associated with this representation. The method
works for dilute systems where all the spins
can be considered independently. Unfortunately, attempts to generalize this technique to the lattice
of spins results in the replacement of the local constraint (the number of particles on each site is
fixed) by the so-called global constraint where the number of particles is fixed only on an average
for the whole crystal. There is no reason to believe that such an approximation is a good starting
point for the description of the strongly correlated systems.
Another possibility to treat the local constraint rigorously is based on
Majorana fermion representation. In this case fermions are ``real'' and corresponding gauge symmetry is
$Z_2$. The difficulty with this representation is mostly related to the physical regularization
of the fluctuations associated with the discrete symmetry group.

An alternative approach for spin Hamiltonians, free from local constraint problem,
has been proposed in the pioneering paper of Popov and Fedotov \cite{popov88a}. Based on the exact
fermionic representation for $S=1/2$ and $S=1$ operators, where the constraint is controlled by
purely imaginary Lagrange multipliers, these authors demonstrated the power and simplification
of the corresponding Matsubara diagram technique. The semi-fermionic representation
(we discuss the meaning of this definition in the course of our paper) used by Popov and Fedotov
is neither fermionic, nor bosonic, but reflects the fundamental Pauli nature of spins.
The goal of this paper is to give a brief introduction  to a semi-fermionic (SF) approach. A reader can
find many useful technical details, discussion of mathematical aspects
of semi-fermionic representation and its application to various problems in the
original papers \cite{popov88a}-\cite{col02}.
However, we reproduce the key steps of important derivations
contained in \cite{kis01},\cite{kis02a} in order to make the reader's job easier.

The manuscript is organized as follows: in Section I, the general
concept of semi-fermions is introduced. We begin with the
construction of the SF formalism for the fully antisymmetric
representation of $SU(N)$ group and the fully symmetric SF
representation of $SU(2)$ group using the imaginary-time
(Matsubara) representation. We show a ``bridge'' between different
representations using the simplest example of $S=1$ in $SU(2)$ and
discuss the SF approach for $SO(4)$ group. Finally, we show how to
work with semi-fermions in real-time formalism and construct the
Schwinger-Keldysh technique for SF. In this section, we will
mostly follow original papers by the author \cite{kiselev00b},
\cite{kis01}. The reader acquainted with semi-fermionic technique
can easily skip this section. In Section II, we illustrate the
applications of SF formalism for various problems of condensed
matter physics, such  as ferromagnetic (FM), antiferromagnetic
(AFM) and resonance valence bond (RVB) instabilities in the
Heisenberg model, Dicke model, large negative - U Hubbard and t-J
models, competition between local and non-local correlations in
Kondo lattices in the vicinity of magnetic and spin glass critical
points, dynamical symmetries in quantum dots, spin chains and
ladders. In the Epilogue, we discuss some open questions and
perspectives. Necessary information about dynamical groups is
contained in Appendix A, the effective models for spin chains are
discussed in Appendix B.
\section{Semi-fermionic representation}
To begin with, we briefly reproduce the arguments contained in the
original paper of Popov and Fedotov.
Let's assume first $S=1/2$. We denote as
$H_\sigma$ the Hamiltonian of spin system. The standard Pauli
matrices can be represented as bilinear combination of Fermi operators as follows:
\begin{equation}
\sigma_j^z\to a_j^\dagger a_j - b_j^\dagger b_j,\;\;\;\;\;\;
\sigma_j^+\to 2 a_j^\dagger b_j,\;\;\;\;\;\;\;
\sigma_j^-\to 2 b_j^\dagger a_j.
\label{rep1}
\end{equation}
on each site $i$ of the lattice. The partition function of the spin problem $Z_\sigma$
is given by
\begin{equation}
Z_{\sigma}= Tr \exp(-\beta \hat H_\sigma)=i^N Tr\exp(-\beta (\hat H_F+i\pi \hat N_F/(2\beta))
\label{pf1}
\end{equation}
where $\hat H_F$ is the operator obtained from $\hat H_\sigma$ by the replacement (\ref{rep1}) and
\begin{equation}
\hat N=\sum_{j=1}^N(a^\dagger_j a_j +b^\dagger_j b_j)
\end{equation}
($N$ is the number of sites in the system and $\beta=1/T$ is inverse temperature). To prove equation
(\ref{pf1})
we note that the trace over the nonphysical states of the $i$-th site vanishes
\begin{equation}
Tr_{unphys}\exp(-\beta (\hat H_F+i\pi \hat N_F/(2\beta))= (-i)^0+(-i)^2=0
\end{equation}
Thus, the identity (\ref{pf1}) holds. The constraint of fixed number of fermions $\hat N_j=1$,
 is achieved by means of the
purely imaginary Lagrange multipliers $\mu=-i\pi/(2\beta)$ playing the role of imaginary chemical
potentials of fermions. As a result, the Green's function
\begin{equation}
G=(i\omega_F-\epsilon)^{-1}
\end{equation}
is expressed in terms of  Matsubara frequencies $\omega_F=2\pi T(n+1/4)$  corresponding
neither Fermi nor Bose statistics.

For $S=1$ we adopt the representation of $\hat H_\sigma$ in terms of the 3-component Fermi field:
\begin{equation}
\sigma^z_j\to a^\dagger_j a_j -b^\dagger_j b_j,\;\;\;\;\;
\sigma^+_j\to \sqrt{2}(a_j^\dagger c_j +c_j^\dagger b_j),\;\;\;\;
\sigma^-_j\to \sqrt{2}(c_j^\dagger a_j +b_j^\dagger c_j).
\label{s1}
\end{equation}
The partition function $Z_\sigma$ is given by
\begin{equation}
Z_{\sigma}=Tr (-\beta \hat H_\sigma)=\left(\frac{i}{\sqrt{3}}\right)^N
Tr\exp(-\beta (\hat H_F+i\pi \hat N_F/(3\beta)).
\end{equation}
It is easy to note that the states with occupation numbers 0 and 3 cancel each other, whereas
states with occupation 1 and 2 are equivalent due to the particle-hole symmetry and thus can be taken
into account on an equal footing by proper normalization of the partition function. As a result,
the Green's function in the
imaginary time representation is expressed in terms of $\omega_F=2\pi T(n+1/3)$
frequencies.

In this section, we show how semi-fermionic (Popov-Fedotov) representation can be derived
using the mapping of partition function of the spin problem onto the corresponding partition function of
the fermionic problem. The cases of arbitrary N (even) for SU(N) groups and arbitrary S for SU(2)
group are discussed.
\subsection{SU(N) group}
We begin with the derivation of SF representation for SU(N) group.
The SU($N$) algebra is determined by the generators obeying the
following commutation relations:
\begin{equation}
[\hat S^\beta_{\alpha, i} \hat S^\rho_{\sigma j}]=
\delta_{ij}(\delta^\rho_\alpha \hat S^\beta_{\sigma i}-\delta^\beta_\sigma
\hat S^\rho_{\alpha i}),
\end{equation}
where $\alpha,\beta=1,...,N$. We adopt the definition of the
Cartan algebra \cite{cartan} of the SU($N$) group
$\{H_\alpha\}=S_\alpha^\alpha$ similar to the one used in \cite{sachdev89a},
noting that the diagonal generators $S_\alpha^\alpha$ are not traceless.
To ensure a vanishing trace, the
diagonal generators should only appear in combinations
\begin{equation}
\sum_{\alpha=1}^{N} s_\alpha S_\alpha^\alpha \quad \mbox{with} \quad
\sum_{\alpha=1}^{N} s_\alpha = 0,
\end{equation}
which effectively reduce the number of independent diagonal generators to
$N-1$ and the total number of SU($N$) generators to $N^2-1$.

In this paper we discuss the representations of SU(N) group
determined by rectangular Young Tableau (YT) (see
\cite{sachdev89a} and \cite{kis01} for details) and mostly
concentrate on two important cases of the fully asymmetric (one
column) YT and the fully symmetric (one row) YT (see Fig.1).

\begin{figure}
\begin{center}
  \epsfxsize36mm \epsfbox{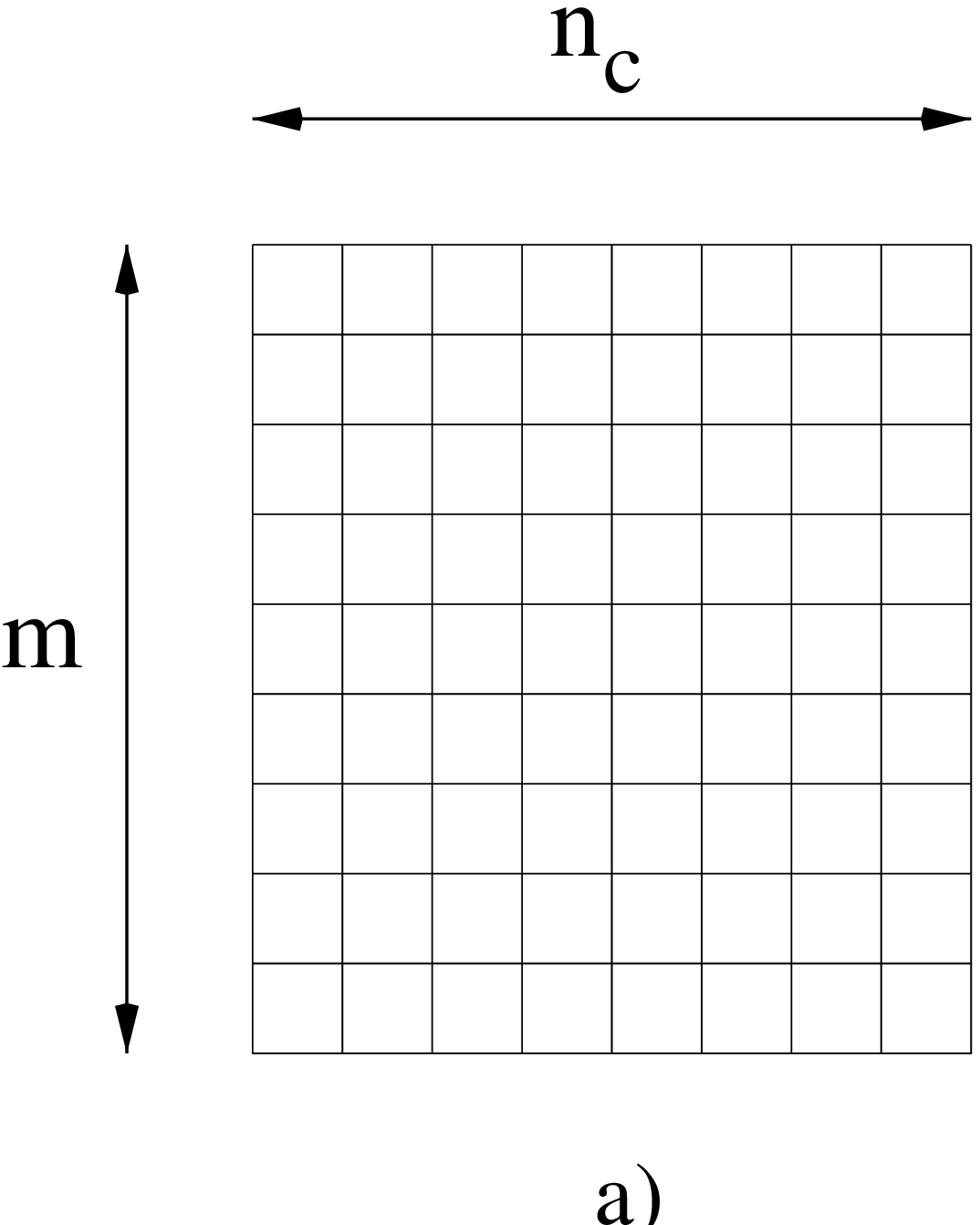}\hspace*{1cm}
  \epsfxsize36mm \epsfbox{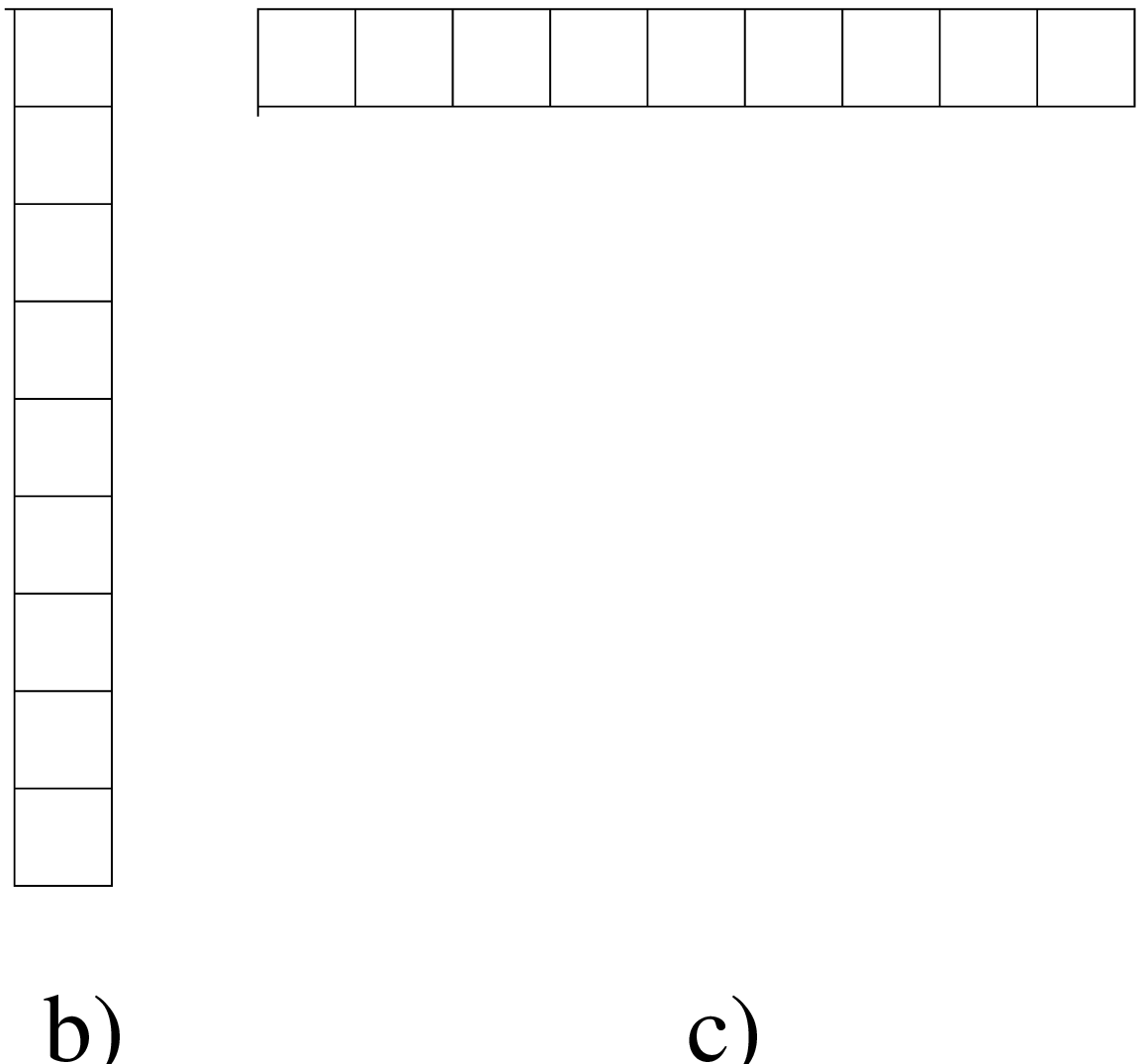}
\end{center}
\caption{ Rectangular  Young tableau to denote a SU($N$)
representation (a), single column tableau corresponding to $n_c=1$
(b) and single row tableau standing for spin $S=n_c/2$
representation of SU(2) group (c).} \label{fig:young_tableaux}
\end{figure}

The generator  $\hat S^\alpha_\beta$  may be written as biquadratic form
in terms of the Fermi-operators
\begin{equation}
\hat S^\alpha_\beta=\sum_\gamma a^\dagger_{\alpha \gamma}a^{\beta \gamma}
\label{fermi}
\end{equation}
where the "color" index $\gamma=1,...,n_c$ and the $n_c(n_c+1)/2$
constraints
\begin{equation}
\sum_{\alpha=1}^Na^\dagger_{\alpha \gamma_1} a^{\alpha \gamma_2}=\delta_{\gamma_{1}}^{\gamma_{2}} m
\label{gconst}
\end{equation}
restrict the Hilbert space to the states with $m * n_c$ particles and
ensure the characteristic symmetry in the color index $a$.  Here $m$ corresponds
to the number of rows in rectangular Young Tableau whereas $n_c$ stands for the number of columns.
The antisymmetric behavior with respect to $\alpha$ is a direct
consequence of the fermionic representation.

Let us consider the partition function for the Hamiltonian, expressed
in terms of SU($N$) generators
\begin{equation}
Z_{S}=Tr \exp(-\beta H_S)= Tr' \exp(-\beta H_F)
\label{z1}
\end{equation}
where $Tr'$ denotes the trace taken with constraints (\ref{gconst}).
As it is shown in \cite{kis01}, the partition function of $SU(N)$ model
is related to partition function of corresponding fermion model through
the following equation:
\begin{equation}
Z_S=\int\prod_{j} d\mu(j) P(\mu(j))
Tr\exp\left(-\beta(H_F-\mu(j)n_F)\right)=
\end{equation}
$$
=\int\prod_{j} d\mu(j) P(\mu(j))Z_F(\mu(j))
$$
here $P(\mu_j)$ is a distribution function of imaginary Lagrange
multipliers. We calculate $P(\mu_j)$ explicitly using constraints
(\ref{gconst}).

We use the path integral representation of the partition function
\begin{equation}
Z_S/Z_S^0=\int\prod_{j} d\mu(j) P(\mu(j))\exp({\cal A})/
\int\prod_{j} d\mu(j) P(\mu(j))\exp({\cal A}_0)
\label{z2}
\end{equation}
where the actions ${\cal A}$ and ${\cal A}_0$ are determined by
\begin{equation}
{\cal A}={\cal A}_0 - \int_0^\beta d\tau H_F(\tau),\;\;\;
{\cal A}_0=\sum_{j}\sum_{k=1}^N\int_0^\beta d\tau\bar a_k(j,\tau)
(\partial_\tau+\mu(j))a_k(j,\tau)
\end{equation}
and the fermionic representation of SU($N$) generators (\ref{fermi})
is applied.

Let us first consider the case $n_c=1$. We denote the corresponding
distribution by $P_{N,m}(\mu(j))$, where $m$ is the number of particles
in the SU($N$) orbital, or in other words, $1\leq m < N$ labels the
different fundamental representations of SU($N$).
\begin{equation}
n_j=\sum_{k=1}^N \bar a_{k}(j) a_{k}(j)=m
\label{const1}
\end{equation}
To satisfy this requirement, the minimal set of chemical
potentials and the corresponding form of $P_{N,m}(\mu(j))$ are to
be derived (see Fig.2).

To derive the distribution function, we use the following identity for the
constraint (\ref{const1}) expressed in terms of Grassmann variables
\begin{equation}
\delta_{n_j, m}=\frac{1}{N}
\sin\left(\pi(n_j-m)\right)/
\sin\left(\frac{\pi(n_j-m)}{N}\right)
\label{d1}
\end{equation}
Substituting this identity into (\ref{z1}) and comparing with
(\ref{z2}) one gets
\begin{equation}
P_{N,m}(\mu(j))=\frac{1}{N}\sum_{k=1}^{N}
\exp\left(\frac{i\pi m}{N}(2k-1)\right)\delta(\mu(j)-\mu_k),
\label{eq:P_v1}
\end{equation}
where
\begin{equation}
\mu_k = - \frac{i \pi T}{N}(2k-1).
\label{eq:mu_k}
\end{equation}
Since the Hamiltonian is symmetric under the exchange of particles and
holes when the sign of the Lagrange multiplier is also changed simultaneously,
we can simplify (\ref{eq:P_v1}) to
\begin{equation}
P_{N,m}(\mu(j))=
\frac{2 i}{N}\sum_{k=1}^{\lfloor N/2 \rfloor}
\sin\left(\pi m\frac{2k-1}{N}\right)\delta(\mu(j)-\mu_k)
\label{dfu}
\end{equation}
where $\lfloor N/2 \rfloor$ denotes the integer part of $N/2$.  As shown below,
this is the minimal representation of the
distribution function corresponding to the minimal set of the discrete
imaginary Lagrange multipliers. Another distributions function
different from (\ref{dfu}) can be constructed when the sum
is taken from $k=N/2+1$ to $N$. Nevertheless, this DF is
different from (\ref{dfu}) only by the sign of imaginary Lagrange multipliers
$\tilde \mu_k=\mu_k^*=-\mu_k$ and thus is supplementary to (\ref{dfu}).

\begin{figure}
\begin{center}
  \epsfysize4cm \epsfbox{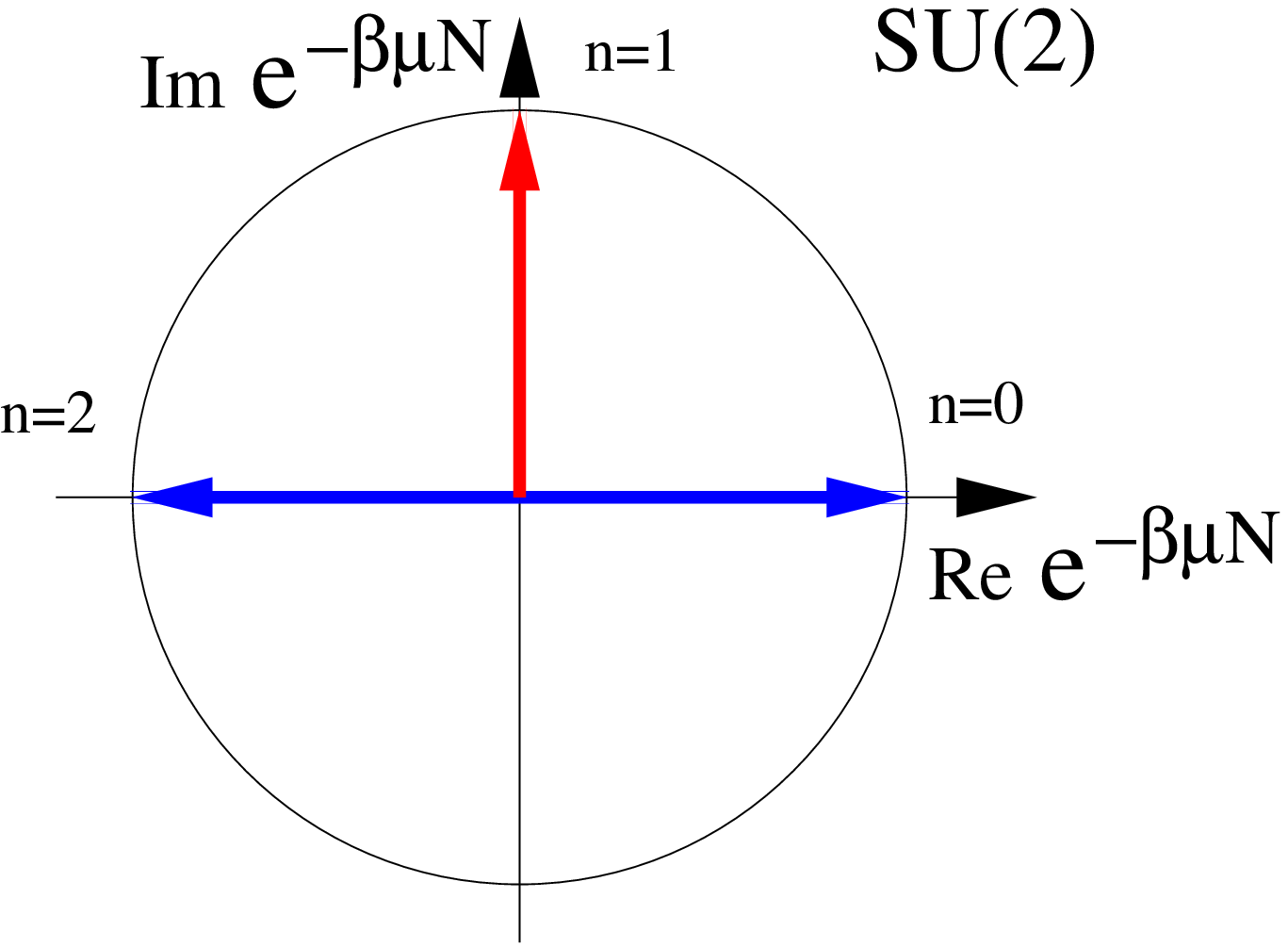}
  \epsfysize4cm
\epsfbox{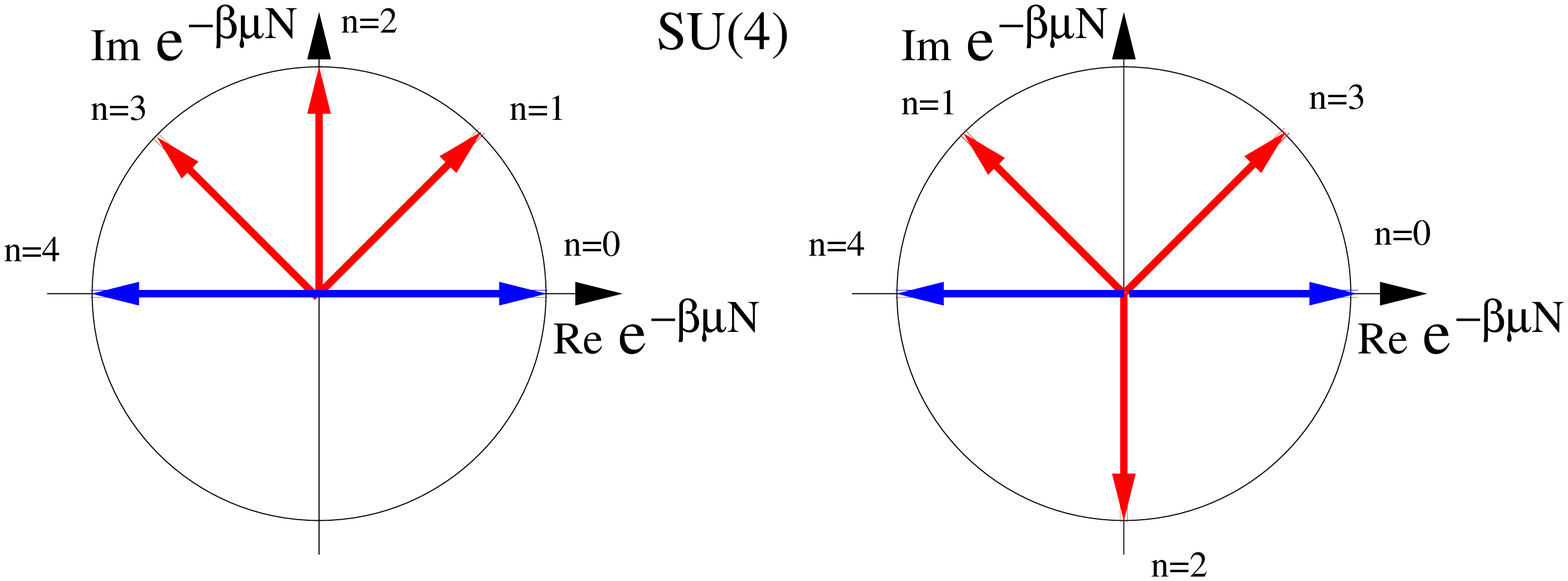}
\end{center}
\caption{Graphical representation of exclusion principle for
SU($N$)
  semi-fermionic representation with even $N$, $n_c=1$
(we use $\mu=i\pi T/2$ for SU(2) and $\mu_1=i\pi T/4,\;\mu_2=3 i
\pi T/4$ for SU(4)).} \label{fig:SU(N)_rep}
\end{figure}

Particularly interesting for even $N$ is the case when the SU($N$)
orbital is half--filled, $m=N/2$.  Then all Lagrange multipliers carry  equal weight
\begin{equation}
P_{N,N/2}(\mu(j))=\frac{2i}{N}\sum_{k=1}^{N/2}(-1)^{k+1}\delta\left(\mu(j)-\mu_k\right).
\label{nn2}
\end{equation}
Taking the limit $N\to\infty$ one may replace the summation in
expression (\ref{nn2}) in a suitable way by integration.  Note, that while
taking $N\to\infty$ and $m\to\infty$ limits, we nevertheless keep the ratio
$m/N=1/2$ fixed. Then, the
following limiting distribution function
can be obtained:
\begin{equation}
P_{N,N/2}(\mu(j)) \stackrel{N\to\infty}{\longrightarrow}
\frac{\beta}{2\pi i}
\exp\left(-\beta \mu(j) \frac{N}{2}\right)
\label{pg}
\end{equation}
resulting in the usual continuous representation of the local
constraint for the simplest case $n_c=1$
\begin{equation}
Z_S=Tr(\exp\left(-\beta H_F\right) \delta \left(n_j
-\frac{N}{2}\right)
\end{equation}
We note the obvious similarity of the limiting DF (\ref{pg}) with the
{\it Gibbs canonical distribution} provided that
the Wick rotation from the imaginary axis of the Lagrange multipliers $\mu$
to the real axis of energies $E$ is performed and thus $\mu(j) N/2$ has a
meaning of energy.

\begin{figure}
\begin{center}
  \epsfxsize8cm \epsfysize5cm \epsfbox{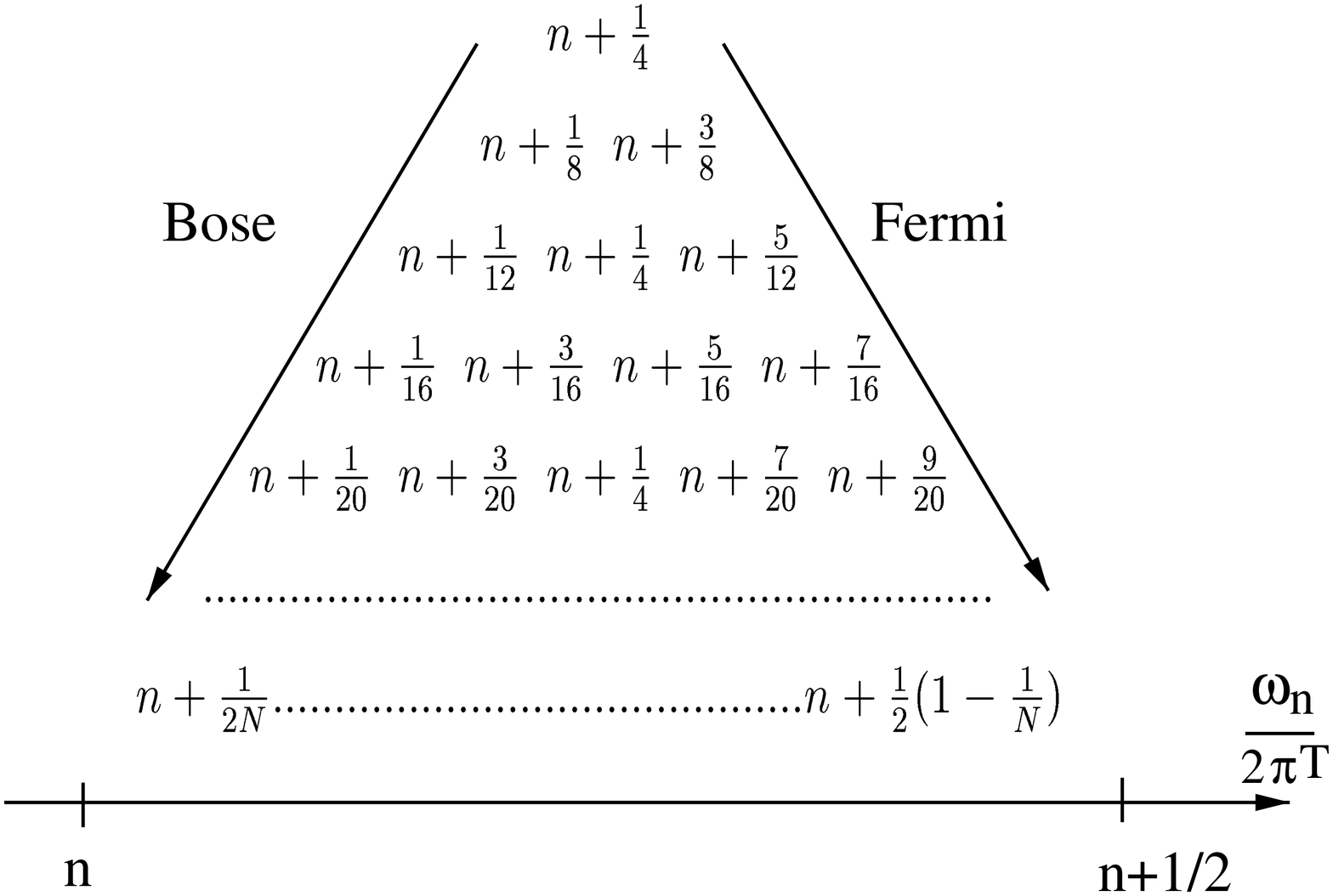}\\
  \epsfxsize8cm \epsfysize5cm \epsfbox{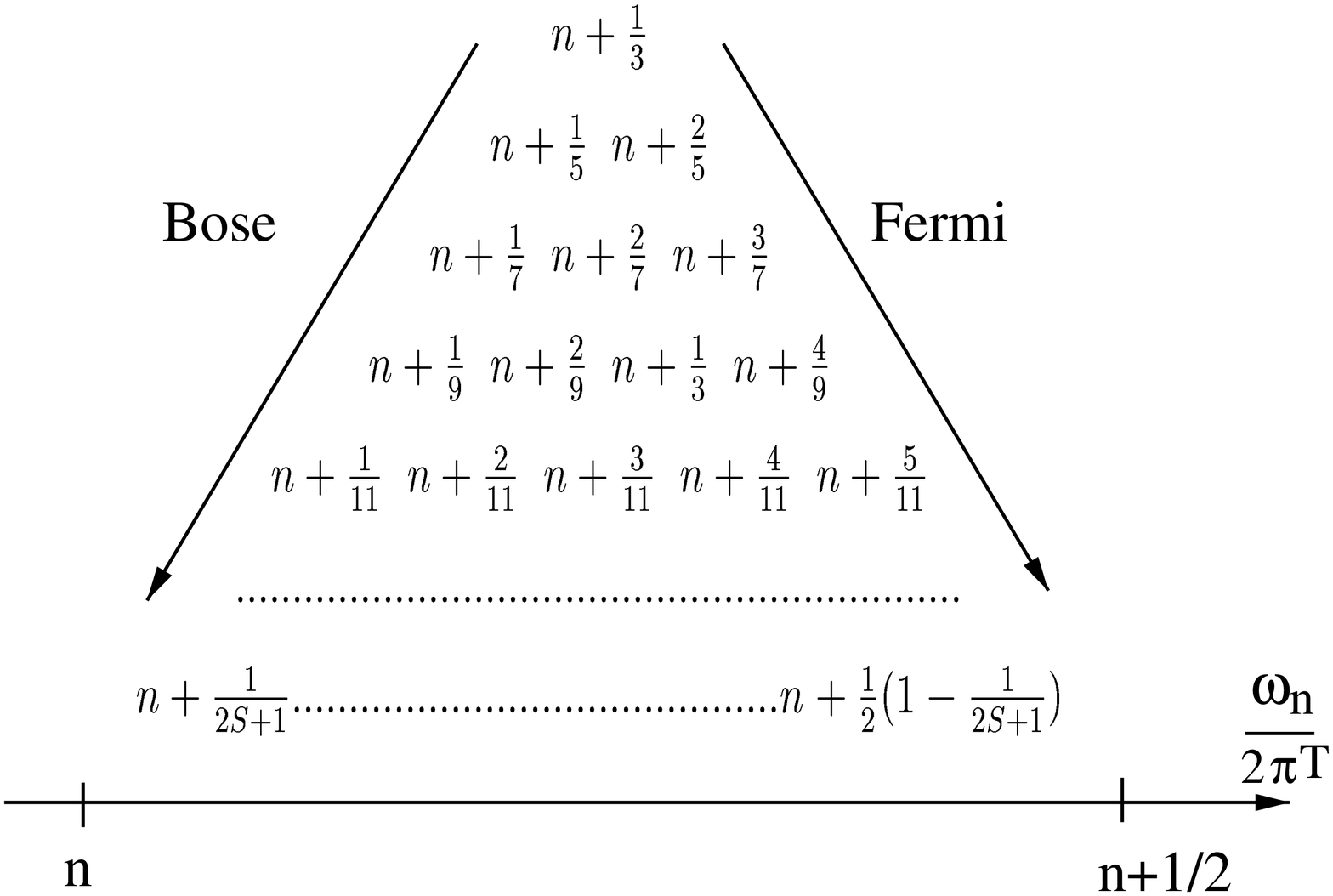}
\end{center}
\caption{The minimal set of Matsubara frequencies for a) $SU(N)$
representation with even $N$/ $SU(2)$ representation for
half-integer value of the spin. b) $SU(2)$  representation for
integer values of the spin and $l=1$.}
\label{fig:frequencies_spin}
\end{figure}

Up to now, the representation we discussed was purely fermionic and
expressed in terms of usual Grassmann variables when the path integral
formalism is applied. The only difference from slave fermionic
approach is that imaginary Lagrange multipliers are introduced to
fulfill the constraint.  Nevertheless, by making the replacement
\begin{equation}
a_k(j,\tau)) \to a_k(j,\tau)
  \exp\left(\frac{i\pi\tau}{\beta} \frac{2k-1}{N}\right),\;\;\;
  \bar a_k(j,\tau) \to \bar a_k(j,\tau)
  \exp\left(-\frac{i\pi\tau}{\beta} \frac{2k-1}{N}\right)
\end{equation}
we arrive at the generalized Grassmann (semi-fermionic) boundary
conditions
\begin{equation}
 a_k(j,\beta) = a_k(j,0)\exp\left(i\pi \frac{2k-1}{N}\right),\;\;\;
          \bar a_k(j,\beta) = \bar a_k(j,0)\exp\left(-i\pi
            \frac{2k-1}{N}\right)
\label{bk}
\end{equation}
This leads to a temperature diagram technique for the Green's functions
\begin{equation}
{\cal G}^{\alpha\beta}(j,\tau)=-
\langle T_\tau a_\alpha(j,\tau) \bar a_\beta(j,0)\rangle
\label{itf}
\end{equation}
of semi-fermions with Matsubara frequencies different from both Fermi
and Bose representations (see Fig.\ref{fig:frequencies_spin}).

The exclusion principle for this case is illustrated on
Fig.\ref{fig:SU(N)_rep}, where the $S=1/2$ representation for the
first two groups SU(2) and SU(4) are shown. The first point to
observe is that the spin Hamiltonian does not distinguish the $n$
particle and the $n$ hole (or $N-n$ particle) subspace. Eq.
(\ref{eq:mu_k}) shows that the two phase factors $\exp(\beta \mu
n)$ and $\exp(\beta \mu (N-n))$ accompanying these subspaces in
Eq. (\ref{dfu}) add up to a purely imaginary value within the same
Lagrange multiplier, and the empty and the fully occupied states
are always cancelled. In the case of $N \geq 4$, where we have
multiple Lagrange multipliers, the distribution function $P(\mu)$
linearly combines these imaginary prefactors to select out the
desired physical subspace with particle number $n=m$.

In Fig.\ref{fig:SU(N)_rep}, we note that on each picture, the
empty and fully occupied states are cancelled in their own unit
circle. For SU(2) there is a unique chemical potential $\mu=\pm
i\pi T/2$ which results in the survival of single occupied states.
For SU(4) there are two chemical potentials (see also
Fig.\ref{fig:frequencies_spin}). The cancellation of single and
triple occupied states is achieved with the help of proper weights
for these states in the distribution function whereas the states
with the occupation number 2 are doubled according to the
expression (\ref{nn2}). In general, for SU($N$) group with $n_c=1$
there exists $N/2$ circles providing the realization of the
exclusion principle.

\subsection{SU(2) group}

We consider now the generalization of the SU(2) algebra for the case of spin $S$.
Here, the most convenient fermionic representation is constructed with
the help of a $2S+1$ component Fermi field $a_k(j)$ provided that the
generators of SU(2) satisfy the following equations:
$$
 S^+=\sum_{k=-S}^{S-1}\sqrt{S(S+1)-k(k+1)}a^\dagger_{k+1}(j)a_k(j),
$$
$$
 S^-=\sum_{k=-S+1}^{S}\sqrt{S(S+1)-k(k-1)}a^\dagger_{k-1}(j)a_k(j),
$$
\begin{equation}
S^z=\sum_{k=-S}^S k a^\dagger_{k}(j)a_k(j)
\label{su2}
\end{equation}
such that $dim H_F=2^{2S+1}$ whereas the constraint reads as follows
\begin{equation}
n_j=\sum_{k=-S}^{k=S}a^\dagger_{k}(j) a_{k}(j)=l=1
\label{const2}
\end{equation}
 Following the same routine as for
SU($N$) generators and using the occupancy condition to have $l=1$ (or
$2S$) states of the $(2S+1)$ states filled, one gets the following
distribution function, after using the particle--hole symmetry of the Hamiltonian
$H_S$:
\begin{equation}
P_{2S+1,1}(\mu(j))=\frac{2i}{2S+1}\sum_{k=1}^{\lfloor S+1/2\rfloor}
\sin\left(\pi \frac{2k-1}{2S+1}\right)\delta(\mu(j)-\mu_k)
\label{lsu}
\end{equation}
where the Lagrange multipliers are $\mu_k=-i\pi T(2k-1)/(2S+1)$ and
$k=1,...,\lfloor S+1/2 \rfloor$, similarly to Eq.(\ref{eq:mu_k}).

In the particular case of the SU(2) model  for some chosen
values of spin $S$ the distribution functions are given by the
following expressions
\begin{equation}
  P_{2,1}(\mu(j))=i\;\delta\left(\mu(j)+\frac{i\pi T}{2}\right)
\nonumber
\end{equation}
for $S=1/2$
\begin{equation}
 P_{3,1}(\mu(j))=P_{3,2}(\mu(j))=
  \frac{i}{\sqrt{3}}\;\delta\left(\mu(j)+\frac{i\pi T}{3}\right)
\nonumber
\end{equation}
for $S=1$.

This result corresponds to the original Popov-Fedotov description
restricted to the $S=1/2$ and $S=1$ cases.
\begin{figure}
\begin{center}
  \epsfysize4cm  \epsfbox{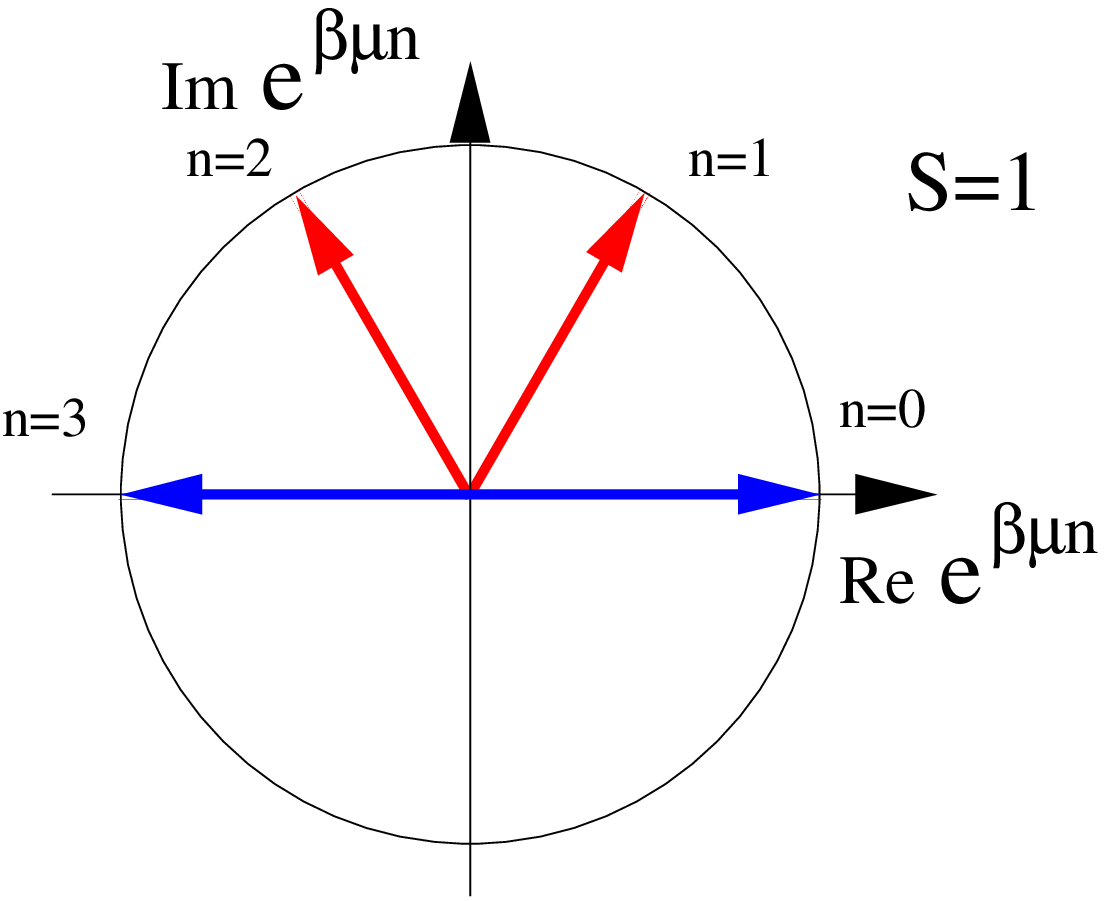}
   \epsfysize4cm \epsfbox{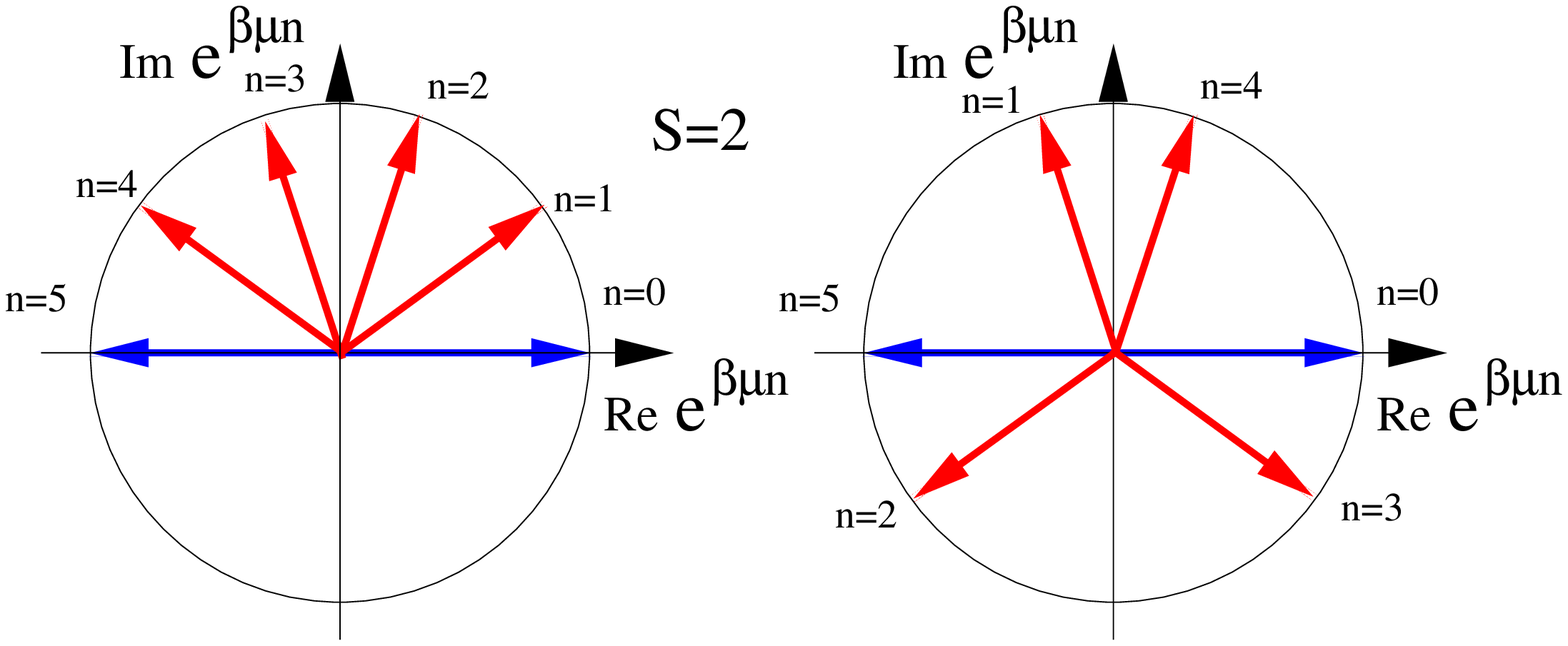}
\end{center}
\caption{Graphical representation of exclusion principle for SU(2)
  semi-fermionic representation for $S=1$  and $S=2$.
  For any arbitrary integer
  value of spin there exists $S$ circle diagrams corresponding to the
  $S$ different chemical potentials and providing the realization of
  the exclusion principle.}
\label{fig:spin_rep}
\end{figure}

We present as an example some other distribution functions
obtained according to general scheme considered above:
$$
 P_{4,1}(\mu)=P_{4,3}(\mu)=
$$
\begin{equation}
 =\frac{i\sqrt{2}}{4}\left(
    \delta(\mu+\frac{i\pi T}{4})+
    \delta(\mu+\frac{3i\pi T}{4})\right)
\nonumber
\end{equation}
for $S=3/2$, SU(2) and
\begin{equation}
  P_{4,2}(\mu)=\frac{i}{2}\left(\delta(\mu+\frac{i\pi T}{4})-
    \delta(\mu+\frac{3i\pi T}{4})\right)
\nonumber
\end{equation}
for effective spin "$S=1/2$", SU(4),

$$
 P_{5,1}(\mu)=P_{5,4}(\mu)=
$$
\begin{equation}\displaystyle
 =\frac{i}{\sqrt{10}}\left(\sqrt{1-\frac{1}{\sqrt{5}}}\;\;
    \delta(\mu+\frac{i\pi T}{5})+
   \sqrt{1+\frac{1}{\sqrt{5}}}\;\; \delta(\mu+\frac{3i\pi T}{5})\right)
\nonumber
\end{equation}
for $S=2$, $SU(2)$ etc.

A limiting distribution function corresponding to Eq. (\ref{pg}) for
the constraint condition with arbitrary $l$ is found to be
\begin{equation}
P_{\infty,l}(\mu(j))  \stackrel{S\to\infty}{\longrightarrow}
\frac{\beta}{2\pi i}\exp(-\beta
l\mu(j)).
\end{equation}
For the  case $l=m=N/2\to\infty$ and
$S=(N-1)/2\to\infty$
the expression for the limiting DF $P_{\infty,l}(\mu(j))$ coincides with (23).
We note that in $S\to\infty$ (or $N\to\infty$) limit, the  continuum
``chemical potentials'' play the role of additional U(1) fluctuating field
whereas for finite $S$ and $N$ they are characterized by fixed and
discrete values.

When $S$ assumes integer values, the minimal fundamental set of
Matsubara frequencies is given by the table in
Fig.\ref{fig:frequencies_spin}.

The exclusion principle for SU(2) in the large spin limit can be
also understood with the help of Fig.\ref{fig:SU(N)_rep} and
Fig.\ref{fig:spin_rep}.  One can see that the empty and the fully
occupied states are cancelled in each given circle similarly to
even-$N$ SU($N$) algebra.  The particle-hole (PH) symmetry of the
representation results in an equivalence of single occupied and
$2S$ occupied states whereas all the other states are cancelled
due to proper weights in the distribution function (\ref{lsu}). In
accordance with PH symmetry being preserved for each value of the
chemical potential all circle diagrams (see Fig.3, Fig.5) are
invariant with respect to simultaneous change $\mu \leftrightarrow
-\mu$ and $n_{particle} \leftrightarrow n_{holes}$.

\subsection{Semi-fermionic representation of $SO(N)$ group}
In this chapter we show the way of generalization of
semi-fermionic representation to the dynamical algebras $o(4)$ and
$o(5)$ (see Appendix A). Like in the case of pure spin operators,
these representations should preserve all kinematical constraints.

The first step to derive semi-fermionic representation for SO(4)
group is based on the local isomorphism of $SO(4)$ and
$SU(2)\times SU(2)$.

We start with $n=4$ - field representation of $SU(2)$ group
(\ref{fermi})
\begin{equation}
(a_{11},~ a_{12},~ a_{21},~ a_{22}) \label{1}
\end{equation}
There are two diagonal and two off-diagonal constraints
(\ref{gconst}) which read as follows:
\begin{equation}
a^\dagger_{11}a_{11}+a^\dagger_{21}a_{21}=1,\;\;\;\;
a^\dagger_{12}a_{12}+a^\dagger_{22}a_{22}=1,
\end{equation}
\begin{equation}
a^\dagger_{11}a_{12}+a^\dagger_{21}a_{22}=0,\;\;\;\;
a^\dagger_{12}a_{11}+a^\dagger_{22}a_{21}=0, \label{so4c1}
\end{equation}
and generators of $SU(2)$ group are given by
$$
S^-=S_2^1=a^\dagger_{11}a_{21}+a^\dagger_{12}a_{22},\;\;\;\;
S^+=S_1^2=a^\dagger_{21}a_{11}+a^\dagger_{22}a_{12}
$$
\begin{equation}
2 S^z=S_2^2-S_1^1=
a^\dagger_{21}a_{21}+a^\dagger_{22}a_{22}-a^\dagger_{11}a_{11}-a^\dagger_{12}a_{12}
\label{sz}
\end{equation}
Combining definition (\ref{sz}) with constraint (\ref{so4c1}) we
reach the  following equations:
\begin{eqnarray}
S^-&=&a^\dagger_{11}(a_{21}+
a_{12})+(a^\dagger_{12}+a^\dagger_{21})a_{22},\nonumber\\
S^+&=&(a^\dagger_{21}+
a^\dagger_{12})a_{11}+a^\dagger_{22}(a_{12}+a_{21}),\nonumber\\
S^z&=&a^\dagger_{22}a_{22}-a^\dagger_{11}a_{11}
\end{eqnarray}
Therefore, we conclude that the antisymmetric (singlet)
combination $a_{12}-a_{21}$ does not enter the expression for spin
$S=1$ operators. Thus, three (out of four) component Fermi-field
is sufficient for the description of $S=1$ $SU(2)$ representation
in agreement with (\ref{su2}). Defining new fields as follows
\begin{equation}
a_{11}=f_{-1},\;\;\;\; a_{22}=f_1,\;\;\;\;\;
\frac{1}{\sqrt{2}}(a_{12}+a_{21})=f_0,\;\;\;\;\;\;
\frac{1}{\sqrt{2}}(a_{12}-a_{21})=f_s. \label{2}
\end{equation}
where fermions $f_1,f_0,f_{-1}$ stand for $S^z=1,0-1$ projections
of the triplet state and fermion $s$ determines the singlet state,
we come to standard $S=1$ $SU(2)$ representation
\begin{equation}
S^+ =  \sqrt{2}(f_0^\dagger f_{-1}+f^\dagger_{1}f_0),\;\;\;\;\;
S^- = \sqrt{2}(f^\dagger_{-1}f_0+ f_0^\dagger f_{1}), \;\;\;\;\;
S_z = f^\dagger_{1}f_{1}-f^\dagger_{-1}f_{-1}, \label{3}
\end{equation}
with the constraint
\begin{equation}
n_1+n_0+n_{-1}+n_s=2 \label{so4c2}
\end{equation}
where $n_\alpha=f^\dagger_\alpha f_\alpha$.

The constraint (\ref{so4c2}) transforms to a standard $SU(2)$
$S=1$ constraint (\ref{const2}) in both cases $n_s=0$ and $n_s=1$
since there is no singlet/triplet mixing allowed by $SU(2)$
algebra.

To demonstrate the transformation of the local constraint let's
first consider the case $n_s=0$. The constraint reads as follows

\begin{equation}
n_1+n_0+n_{-1}=2S\;\;\;\;\;\iff\;\;\;\;\;\;\;{\bf S}^2=S(S+1).
\end{equation}
On the other hand, the states with $2S$ occupation are equivalent
to the states with single occupation due to particle-hole
symmetry. Thus, the constraint (\ref{so4c2}) might be written as
\begin{equation}
\tilde n_1+\tilde n_0+\tilde n_{-1}=1
\end{equation}
where $\tilde n_\alpha =1-n_\alpha$. The latter case corresponds
to $n_s=1$.

The singlet/triplet mixing is allowed for $SO(4)$ group. This
mixing is described in terms of 3 additional generators
responsible for transitions between singlet and triplet (see
Appendix A). Using 4-component auxiliary fermions $f_\lambda$,
where $\lambda=-1,0,1,s$ and matrix form of generators
(\ref{matrix}) we represent $R$ - generators of $SO(4)$ as follows
\begin{eqnarray}
R^+  &=& \sqrt{2}(f^\dagger_{1} f_s -  f_s^\dagger f_{-1}),
\;\;\;\; R^- = \sqrt{2}(f_s^\dagger f_{1} - f^\dagger_{-1}f_s)
,\;\;\;\; R^z = -( f_0^\dagger f_s + f_s^\dagger f_0).
\label{4}\nonumber
\end{eqnarray}
with the only constraint
\begin{equation}
n_1+n_0+n_{-1}+n_s=1,\;\;\;\;\;\iff\;\;\;\;\;\;\;{\bf S}^2+{\bf
R}^2=3 \label{const}
\end{equation}
whereas the orthogonality condition ${\bf S}\cdot{\bf R}={\bf
R}\cdot{\bf S}=0$ is fulfilled automatically. The constraint
(\ref{const}) is respected by means of introducing real chemical
potential $\lambda \to \infty$ for Abrikosov's auxiliary fermions
or imaginary chemical potentials $\mu_t= -i\pi T/3$ for
Popov-Fedotov semi-fermion.

The fermionic representation of $SO(5)$ group is easily
constructed by use semi-fermionic representation and is
characterized by 5-vector ${\bf q}^T=(f_{-1}^\dagger f_0^\dagger,
f_1^\dagger, f_s^\dagger, f_r^\dagger)$
\begin{eqnarray}
S^+ &=&  \sqrt{2}(f_0^\dagger f_{-1}+f^\dagger_{1}f_0),\;\;\;\;
S^z =
f^\dagger_{1}f_{1}-f^\dagger_{-1}f_{-1},\nonumber \\
R^+  &=& \sqrt{2}(f^\dagger_{1} f_s -  f_s^\dagger f_{-1}),
\;\;\;\; R^z = -( f_0^\dagger f_s + f_s^\dagger f_0),\nonumber\\
P^+  &=& \sqrt{2}(f^\dagger_{1} f_r -  f_r^\dagger f_{-1}),
\;\;\;\;  P^z = -( f_0^\dagger f_r + f_r^\dagger f_0). \label{fff}
\end{eqnarray}
and
$$
A=i(f_r^\dagger f_s - f_s^\dagger f_r)
$$
The constraint
\begin{equation}
n_1+n_0+n_{-1}+n_s+n_r=1 \label{const2}
\end{equation}
is respected either by real infinite chemical potential (Abrikosov
pseudofermions) or by set of complex chemical potentials
(semi-fermions). We do not present here ten $5\times 5$ matrices
characterizing $SO(5)$ representation to save a space. The reader
can easily construct them using representations (\ref{fff}). There
exists also a bosonic representation based on Schwinger bosons
which might be derived by the method similar to used above for
$SO(4)$ group. The representations of higher $SO(n)$ groups can be
constructed in a similar fashion.

Kinematic constraints imposed on auxiliary fermions and bosons is
in strict compliance with the Casimir and orthogonality
constraints in spin space. Accordingly, the number of fermionic
and bosonic fields reproduces the dimensionality of spin space
reduced by these constraints. We have seen that the 6-D space of
generators of $SO(4)$ group is reduced to D=4. Then the minimal
(unconstraint) fermionic representation for this group should
contain two $U(1)$ fermions. This means that the representation
(\ref{4}) is not minimal. Apparently, the best way to find such
representation is to use Jordan-Wigner-like transformation
\cite{book2}. The exact form of Jordan-Wigner transformation for
$SO(4)$ group was recently reported in \cite{kis05}. The same kind
of arguments applied to $SO(5)$ group tells us that the spinor
field should contain seven components. This means that the 3-color
U(1) fermionic representation  should be completed by one more
real (Majorana) fermion, and this fact points to one more hidden
$Z_2$ symmetry \cite{book2}.

\subsection{Real-time formalism}
We discuss finally the real-time formalism based on the semi-fermionic
representation of SU($N$) generators. This approach is necessary for
treating the systems out of equilibrium, especially for many
component systems describing Fermi (Bose) quasiparticles interacting
with spins. The real time formalism \cite{keldysh}, \cite{schwinger} provides an alternative approach
for the analytical continuation method for equilibrium problems
allowing direct calculations of correlators whose analytical
properties as function of many complex arguments can be quite
cumbersome.

\begin{figure}
\begin{center}
  \epsfxsize8cm \epsfbox{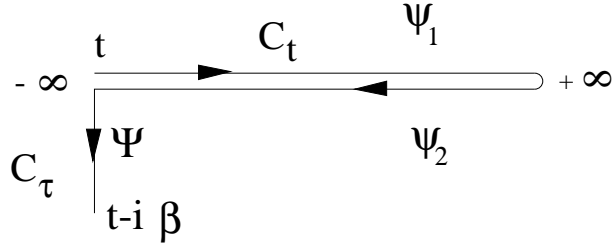} \vspace*{5mm}
\end{center}
\caption{The Keldysh contour going from $-\infty\to\infty\to-\infty$ in real
  time. The boundary conditions on the imaginary time segment
  determine the generalized distribution functions for
  quasiparticles.}
\label{fig:keldysh}
\end{figure}

To derive the real-time formalism for SU($N$) generators we use the
path integral representation along the closed time Keldysh contour
(see Fig.\ref{fig:keldysh}).
Following the standard route \cite{babichenko86a}, we can express the
partition function of the problem containing SU($N$) generators as a
path integral over Grassmann variables
$\psi_l=(a_{l,1}(j),...,a_{l,N}(j))^{T}$ where $l=1,2$ stands for
upper and lower parts of the Keldysh contour, respectively,
\begin{equation}
{\cal Z}/{\cal Z}_0=\int D\bar\psi D\psi\exp(i{\cal A})/
\int D\bar\psi D\psi\exp(i{\cal A}_0)
\label{pfunk}
\end{equation}
where the actions ${\cal A}$ and ${\cal A}_0$ are taken as an integral
along the closed-time contour $C_t+C_\tau$ which is shown in Fig.\ref{fig:keldysh}.
The contour is closed at $t=-\infty+i\tau$ since $\exp(-\beta
H_0)=T_\tau\exp\left(-\int_0^\beta H_0 d\tau\right).$ We denote the
$\psi$ fields on upper and lower sides of the contour $C_t$ as
$\psi_1$ and $\psi_2$ respectively. The fields $\Psi$ stand for the
contour $C_\tau$. These fields provide the matching conditions for
$\psi_{1,2}$ and are excluded from the final expressions.  Taking into
account the semi-fermionic boundary conditions for generalized
Grassmann fields (\ref{bk}) one gets the matching conditions for
$\psi_{1,2}$ at $t=\pm\infty$,
$$
\psi^\mu_{1,\alpha}|_k(-\infty) =
  \exp\left(i\pi\frac{2k-1}{N}\right)\psi^\mu_{2,\alpha}|_k(-\infty),
$$
\begin{equation}
  \psi^\mu_{1,\alpha}|_k(+\infty) =\psi^\mu_{2,\alpha}|_k(+\infty)
\end{equation}
for $k=1,...,\lfloor N/2 \rfloor$ and $\alpha=1,...,N$.  The
correlation functions can be represented as functional derivatives of
the generating functional
\begin{equation}
Z[\eta]={\cal Z}_0^{-1}\int D\bar\psi D\psi\exp\left(i{\cal A}+
i\oint_C d t (\bar\eta\sigma^z\psi+\bar\psi\sigma^z \eta)\right)
\end{equation}
where $\eta$ represents sources and the $\sigma^z$ matrix stands for
"causal" and "anti-causal" orderings along the contour.

The on-site Green's functions (GF) which are matrices of size
$2N\times 2N$ with respect to both Keldysh (lower) and spin-color
(upper) indices are given by
\begin{equation}
G_{\mu\nu}^{\alpha\beta}(t,t')=
-i\frac{\delta}{i\delta\bar \eta_\mu^\alpha(t)}
\frac{\delta}{i\delta \eta_\nu^\beta(t')}
Z[\eta]|_{\bar\eta,\eta\to 0}.
\label{rtf}
\end{equation}
To distinguish between imaginary-time (\ref{itf}) and
real-time (\ref{rtf}) GF's,
we use different notations for Green's functions in these representations.

After a standard shift-transformation \cite{babichenko86a} of the fields $\psi$
the Keldysh GF of free semi-fermions assumes the form
\begin{eqnarray}
G_0^\alpha(\epsilon)=G^{R,\alpha}_0
\left(
\begin{array}{cc}
1 - f_\epsilon &  -f_\epsilon\\
1 - f_\epsilon &  -f_\epsilon
\end{array}\right)-
G^{A,\alpha}_0
\left(
\begin{array}{cc}
-f_\epsilon & -f_\epsilon\\
1 - f_\epsilon & 1 - f_\epsilon
\end{array}
\right),
\nonumber
\end{eqnarray}
where the retarded and advanced GF's are
\begin{equation}
G^{(R,A)\alpha}_0(\epsilon)=(\epsilon \pm i\delta)^{-1},
\quad
f_\epsilon=f^{(N,k)}(\epsilon),
\end{equation}
with equilibrium distribution functions
\begin{equation}
f^{(N,k)}(\epsilon)=T\sum_n\frac{e^{i\omega_{n_k}\tau|_{+0}}}
{i\omega_{n_k}-\epsilon}=
\frac{1}{e^{i\pi (2k-1)/N}\exp(\beta\epsilon)+ 1}.
\end{equation}
A straightforward calculation of $f^{(N,k)}$ for the case of even $N$
leads to the following expression
\begin{equation}
f^{(N,k)}(\epsilon)
=\frac{\displaystyle\sum_{l=1}^N(-1)^{l-1}
\exp\left(\beta\epsilon(N-l)\right)
\exp\left(-\frac{i\pi l(2k-1)}{N}\right)}{\exp(N\beta\epsilon)+1},
\end{equation}
where $k=1,...,N/2.$
The equilibrium distribution functions (EDF) $f^{(2S+1,k)}$ for the
auxiliary Fermi-fields representing arbitrary $S$ for $SU(2)$ algebra
are given by
\begin{equation}
f^{(2S+1,k)}(\epsilon)
=\frac{\displaystyle\sum_{l=1}^{2S+1}(-1)^{l-1}
\exp\left(\beta\epsilon(2S+1-l)\right)
\exp\left(-\frac{i\pi(2k-1)}{2S+1})\right)}
{\exp((2S+1)\beta\epsilon)+(-1)^{2S+1}}
\end{equation}
for $k=1,...,\lfloor S+1/2 \rfloor$.  Particularly simple are the
cases of $S=1/2$ and $S=1$,
\begin{equation}
 f^{(2,1)}(\epsilon)=
    n_F(2\epsilon)-i\frac{1}{2\cosh(\beta\epsilon)}
\label{complex1}
\end{equation}
\begin{equation}
     f^{(3,1)}(\epsilon)=
    \frac{1}{2}n_B(\epsilon)-\frac{3}{2}n_B(3\epsilon)
    -i\sqrt{3}\frac{\sinh(\beta\epsilon/2)}{\sinh(3\beta\epsilon/2)}
\end{equation}
Here, the standard notations for Fermi/Bose distribution functions
$n_{F/B}(\epsilon)=[\exp(\beta\epsilon) \pm 1]^{-1}$ are used. For $S=1/2$ the
semi-fermionic EDF satisfies the
obvious identity $|f^{(2,1)}(\epsilon)|^2=n_F(2\epsilon)$.

In general the EDF for half-integer and integer spins can be expressed in
terms of Fermi and Bose EDF respectively.  We note that since
auxiliary Fermi fields introduced for the representation of SU($N$)
generators do not represent the true quasiparticles of the problem,
helping only to treat properly the constraint condition, the
distribution functions for these objects in general do not have to be
real functions.  Nevertheless, one can prove that the imaginary part
of the EDF does not affect the physical correlators and can be
eliminated by introducing an infinitesimally small real part for the
chemical potential.  In spin problems, a uniform/staggered magnetic
field usually plays the role of such real chemical potential for
semi-fermions.

\section{Application of semi-fermionic representation for strongly correlated systems}
In this Section we illustrate some of the applications of SF
representation for various problems of strongly correlated
physics.

\subsection{Heisenberg model: FM, AFM and RVB}

The effective nonpolynomial action \cite{kol1,kol2} for Heisenberg
model with ferromagnetic (FM) coupling has been investigated in
\cite{popov88a}. The model with antiferromagnetic (AFM)
interaction has been considered by means of semi-fermionic
representation in \cite{kiselev99a} and \cite{azakov} (magnon
spectra) and in \cite{kiselev00b} for resonance valence bond (RVB)
excitations. The Hamiltonian considered is given as
\begin{equation}
H_{int}=-\sum_{<ij>}J_{ij}\left(\vec{S}_i\vec{S}_j-\frac{1}{4}\right)
\label{hh1}
\end{equation}
\begin{itemize}
\item Ferromagnetic coupling $J=I_{FM}>0$
\end{itemize}
The exchange  $\vec{S}_i\vec{S}_j$ is represented as
four-semi-fermion interaction. Applying the Hubbard-Stratonovich
transformation by the {\it local vector} field $\vec{\Phi}_i(\tau)$ the effective nonpolynomial
action is obtained in terms of vector c-field. The FM phase transition corresponds to the
appearance at $T\leq T_c$ of the nonzero average $\langle \Phi^z(q=0,0)\rangle$ which stands for
the nonzero magnetization, or in other words, corresponds to the Bose condensation
of the field $\Phi^z$.
\begin{equation}
\Phi^z(\vec{k},\omega)={\cal
M}(\beta N)^{1/2}\delta_{\vec{k},0} \delta_{\omega,0}
+\tilde\Phi^z(\vec{k},\omega).
\end{equation}
In one loop approximation the
standard molecular field equation can be reproduced
\begin{equation}
{\cal M}=I_{FM}(0)\tanh(\beta{\cal M}/2).
\label{n1}
\end{equation}
The saddle point (mean-field) effective action is given by
well-known expression
\begin{equation}
{\cal A}_0[{\cal M}]=-N\left[\frac{\beta {\cal M}^2}{4I_M(0)}-
\ln\left(2\cosh\left(\frac{\beta {\cal M}}{2}\right)\right)\right],
\end{equation}
and the free energy per spin $f_0$  is determined by the standard
equation:
\begin{equation}
\beta f_0 =-\ln Z_S=\frac{\beta {\cal M}^2}{4I_M(0)}-
\ln\left(2\cosh\left(\frac{\beta {\cal M}}{2}\right)\right)
\end{equation}
Calculation of the second variation of ${\cal A}_{eff}$ gives rise to
the following expression
$$\delta{\cal A}_{eff}=-\frac{1}{4}\sum_{\vec{k}}\Phi^z(\vec{k},0)
\left[I_M^{-1}(\vec{k})-\frac{\beta}{2\cosh^2(\beta\Omega)}\right]
\Phi^z(\vec{k},0)-$$
$$
-\frac{1}{4}\sum_{\vec{k},\omega\ne 0}
I_M^{-1}(\vec{k})\Phi^z(\vec{k},\omega)\Phi^z(\vec{k},\omega)-$$
\begin{equation}
-\sum_{\vec{k},\omega}\Phi^+(\vec{k},\omega)\left[
I_M^{-1}(\vec{k})-\frac{\tanh(\beta\Omega)}{2\Omega - i\omega}\right]
\Phi^-(\vec{k},\omega)
\end{equation}
where $\Omega=(g\mu_B H +{\cal M})/2$. The magnon spectrum ($T\leq T_c$) is
determined by the poles of $\langle \Phi^+\Phi^-\rangle$ correlator, $\omega=\lambda {\bf k}^2$.
\begin{itemize}
\item Antiferromagnetic coupling $J=I_{AFM}<0$. {\it N\'eel solution}
\end{itemize}
The AFM transition corresponds to formation of the staggered condensate
\begin{equation}
\Phi^z(\vec{k},\omega)={\cal N}(\beta N)^{1/2}\delta_{\vec{k},\vec{Q}} \delta_{\omega,0}
+\tilde\Phi^z(\vec{k},\omega)
\end{equation}
The one-loop approximation leads to standard mean-field equations for the staggered magnetization
$$
{\cal N}=-I_{AFM}(Q)\tanh(\beta{\cal N}/2),
$$
\begin{equation}
{\cal A}_0[{\cal N}]=N\left[\frac{\beta {\cal N}^2}{4I_{AFM}(Q)}+
\ln\left(2 \cosh\left( \frac{\beta {\cal N}}{2}\right)\right)\right].
\end{equation}

After taking into account the second variation of ${\cal A}_{eff}$,
the following expression for the effective action is obtained [(see e.g.
\cite{kiselev99a},\cite{azakov}):
$$\delta{\cal A}_{eff}=\frac{1}{4}\sum_{\vec{k}}\Phi^z(\vec{k},0)
\left[I_{AFM}^{-1}(\vec{k})+\frac{\beta}{2\cosh^2(\beta\tilde\Omega)}\right]
\Phi^z(\vec{k},0)+$$
$$
+\frac{1}{4}\sum_{\vec{k},\omega\ne 0}
I_{AFM}^{-1}(\vec{k})\Phi^z(\vec{k},\omega)\Phi^z(\vec{k},\omega)+$$
$$
+\sum_{\vec{k},\omega}\Phi^+(\vec{k},\omega)\left[
I_{AFM}^{-1}(\vec{k})+\frac{2\tilde\Omega\tanh(\beta\tilde\Omega)}
{4\tilde\Omega^2 +\omega^2}
\right]\Phi^-(\vec{k},\omega)-
$$
\begin{equation}
-\sum_{\vec{k},\omega}\Phi^+(\vec{k}+\vec{Q},\omega)
\frac{i\omega}{4\tilde\Omega^2 +\omega^2}
\Phi^-(\vec{k},\omega).
\end{equation}
The AFM magnon spectrum $\omega= c|{\bf k}|$.
\begin{itemize}
\item Antiferromagnetic coupling. {\it Resonance Valence Bond solution}
\end{itemize}
The four-semi-fermion term in (\ref{hh1}) is decoupled by {\it
bilocal scalar} field $\Lambda_{ij}$. The RVB spin liquid (SL)
instability in 2D Heisenberg model corresponds to
Bose-condensation of exciton-like pairs of semi-fermions (for
simplicity we consider the {\it uniform} RVB state):
\begin{equation}
\Delta_0=-\sum_{\bf q}\frac{I_{\bf q}}{I_0}\tanh\left(\frac{I_{\bf q}\Delta_0}{T}\right),
\label{rvb}
\end{equation}
$$
{\cal A}_0=\frac{\beta |I|\Delta_0^2}{2}-\sum_{\bf q}\ln\left[2 \cosh(\beta I_{\bf q}\Delta_0)\right]
$$
where $\Delta_0=\Delta({\bf q}=0)$ is determined by the modulus of $\Lambda_{ij}$ field
\begin{equation}
\Lambda_{<ij>}(\vec{R},\;\vec{r})=
\Delta(\vec{r})\exp\left(i\vec{r}\vec{A}(\vec{R})\right)
\label{uni}
\end{equation}
whereas the second variation of $\delta {\cal A}_{eff}$ describes  the fluctuations of phase $\Lambda_{ij}$
$$
{\cal A}_{eff}=\sum_{{\bf k},\omega}
A_\alpha({\bf k},\omega)\pi^{\alpha\beta}_{{\bf k},\omega}A_\beta({\bf k},\omega),
$$
\begin{equation}
{\Large \pi}^{\alpha\beta}_{{\bf k},\omega}=Tr(p^\alpha p^\beta
(G_{p+k}G_p+G_{p+k}G_p)+\delta_{\alpha\beta}f(I_{\bf p}\Delta_0))
\end{equation}
The spectrum of excitation in uniform SL is determined by zeros of
$\pi^R$ and is purely diffusive \cite{ioffe}-\cite{lee}. The
recent development of application of semi-fermions to low
dimensional magnetic systems can be found in \cite{jean1,jean2}.
\subsection{Dicke model}
In this Section we describe the application of semi-fermionic
approach to two-level systems interacting with single-mode
radiation field (Dicke model). The influence of dissipative
environment on two-level system has been extensively studied in
spin-boson model \cite{leggett} (see also a review \cite{brandes}
for coherent effects in mesoscopic few level system, Dicke super
and sub-radiance effects). Addressing the reader to above
mentioned reviews for discussion of physical implementation of the
Dicke model, we discuss in this section only technical aspects
related to derivation of its equilibrium (thermodynamical)
properties. We closely follow original derivation contained in
Popov-Fedotov paper \cite{popov88a}.

The Dicke Hamiltonian
\begin{equation}
H_D=\omega_0 \psi^\dagger\psi +\frac{\Omega}{2}\sum_i^N\sigma_i^z
+\frac{g}{N^{1/2}}\sum_i^N\left(\sigma_i^+\psi + \psi^\dagger
\sigma_i^-\right)
\end{equation}
contains $\sigma$-matrices representing two-level systems and
bosonic $\psi$-operators describing the single-mode radiation
field.

Applying semi-fermionic representation for two-level system one
gets
\begin{equation}
H_D=\omega_0 \psi^\dagger\psi
+\frac{\Omega}{2}\sum_i^N\left(a_i^\dagger a_i -b_i^\dagger
b_i\right)+\frac{g}{N^{1/2}}\sum_i^N\left(a^\dagger_i b_i \psi +
\psi^\dagger b^\dagger_i a_i\right)
\end{equation}
where $a_i$ and $b_i$ stand for semi-fermionic fields with
generalized Grassmann boundary conditions.

The semi-fermionic variables  appear quadratically in the action
and can be integrated out. As a result, the partition function is
represented as a ratio of two path integrals
\begin{equation}
Z_0/Z_{0\sigma}=\int D[\psi]\exp(S_0[\psi]) (\det
M([\psi]))^N/\int D[\psi]\exp(S_0[\psi]) (\det M([0]))^N
\label{dick1}
\end{equation}
where
$$
S_0[\psi]=\sum_\omega\left(i\omega
-\omega_0\right)\psi^*(\omega)\psi(\omega)
$$
and $M$ is an operator with elements
\begin{eqnarray}
M_{pq}=\left(
\begin{array}{cc}
(ip+\Omega/2)\delta_{pq} & g /(\beta N)^{1/2}\psi^*(p-q)\\
g /(\beta N)^{1/2}\psi(q-p) & (ip-\Omega/2)\delta_{pq}
\end{array}
\right). \end{eqnarray} Here $\omega=2\pi n T$ is bosonic and
$p=2\pi T(n+1/4)$ is semi-fermionic Matsubara frequencies.

Evaluating the integrals (\ref{dick1}) on gets the following
asymptotic for partition function:
\begin{equation}
Z_0/Z_{0\sigma}=\prod_\omega\left[\frac{1}{1-a(\omega)}\right]+O(1/N),\;\;\;\;\,T>T_c
\end{equation}
\begin{equation}
Z_0/Z_{0\sigma}=AN^{1/2}\exp(B N)\prod_\omega
L^{-1}(\omega)+O(1/N^{1/2}),\;\;\;\;\,T<T_c
\end{equation}
where $T_c$ is determined from the equation
\begin{equation}
\frac{g^2}{\omega_0\Omega}\tanh\left(\frac{\Omega}{2T_c}\right)=1
\label{gap1} \end{equation} and the following notations are used:
\begin{equation}
a(\omega)=\frac{g^2\tanh(\Omega/2T)}{(\omega_0-i\omega)(\Omega-i\omega)}
\end{equation}
\begin{equation}
A=\left[\frac{\pi\beta\omega_0\Omega_\Delta^2}{g^2(1-\beta\Omega_\Delta/\sinh(\beta\Omega_\Delta))}\right]^{1/2},\;\;\;\;\;\;
L(\omega)=1+\frac{\omega_0\Omega\omega^2-\omega_0^2\Omega^2}{(\omega^2+\omega_0^2)(\omega^2+\Omega_\Delta^2)},
\end{equation}
\begin{equation}
B=\ln\left[\cosh(\beta\Omega_\Delta/2)/\cosh(\beta\Omega_\Delta/2)\right]-\omega_0\Delta^2\beta,\;\;\;\;\;\;\;\;\;\;
\Omega_\Delta^2=\Omega^2+4g^2\Delta^2.
\end{equation}
The parameter $\Delta(T)$ is determined from Eq.\ref{gap1} with
the replacement $\Omega\to\Omega_\Delta$ and $T_c\to T$.

The Bose spectra are obtained by analytic continuation $i\omega\to
E+i\delta$ of the equations
\begin{equation}
1-a(\omega)=0,\;\;\;\;\;\;T>T_c\;\;\;\;\;\;\;\;\;\;\;\;
L(\omega)=0,\;\;\;\;\;\;T<T_c
\end{equation}
and are given by
\begin{equation}
E_{1,2}=\frac{1}{2}\left[\Omega+\omega_0\mp\sqrt{(\omega-\omega_0)^2+4g^2\tanh(\beta\Omega/2)}\right],\;\;\;\;\;\;T>T_c
\end{equation}
\begin{equation}
E_1=0,\;\;\;\;\,\;\;\;\;E_2=[\left[(\Omega+\omega_0)^2+4g^2\Delta^2\right]^{1/2},\;\;\;\;\;\;T<T_c.
\end{equation}
Multimode variants of the Dicke type and also Dicke models with
interaction that takes non-resonance term into account can be
investigated analogously.

\subsection{Spin-dependent semi-fermionic representation for Hubbard and t-J models}
In this Section we discuss the spin-dependent Popov-Fedotov
representation and application of semi-fermions to Hubbard and t-J
models.

We start with  the negative - U Hubbard model described by the
Hamiltonian
\begin{equation}
H=-\sum_{\langle i,j\rangle,\sigma}t_{ij}a_{i,\sigma}^\dagger
a_{i,\sigma} + U\sum_i n_{i\uparrow}n_{i\downarrow}.
\end{equation}
The interaction $U<0$ is attractive and the hopping amplitude is
assumed small $|t|\ll |U|$. The physical situation underlining
this limit of model  is characterized  by the existence of exactly
one single occupied site and empty or double occupied sites
otherwise. The physical restriction also imposes a subtle
constraint on the corresponding Hilbert space, which requires that
all contributions from states with more than one unpaired electron
are ruled out. Thus, the situation is opposite the Popov-Fedotov
limit where only single occupied states represent physical states
of the model, while empty and double occupied states were
unphysical and eliminated by the single occupancy condition.

In order to remove single occupied states in large negative-U
Hubbard model (LNU) from thermodynamic averages, it was proposed
\cite{oppermann91a,stein92} to introduce an imaginary magnetic
field to the Hamiltonian
\begin{equation}
\tilde H = H- i\alpha\pi T\sum_i\left(n_{i\uparrow} -
n_{i\downarrow}\right)
\end{equation}
with $\alpha$ as a real parameter ($-1\leq \alpha \leq 1$). The
purely imaginary magnetic field term does not violate
time-reversal symmetry \cite{oppermann91a}.

The local partition function $Z^{(i)}=Tr_i\exp(-\beta \tilde H)$
is given by
\begin{equation}
Z^{(i)}=\langle 1,1|e^{-\beta H}|1,1\rangle + \langle
0,0|e^{-\beta H}|0,0\rangle +\gamma \langle 1,0|e^{-\beta
H}|1,0\rangle
\end{equation}
where symmetry condition
$$
\langle 1,0| H|1,0\rangle = \langle 0,1| H|0,1\rangle $$ is used
and $\gamma =2\cos(\pi\alpha)$ is a statistical parameter. The
full fermionic Hilbert space is recovered if $\alpha=0$
($\gamma=2$). The case $\alpha=\pm 1/2$ ($\gamma=0$) result in the
semi-fermionic theory for the LNU-Hubbard model.

If one employs the standard Matsubara temperature diagram
technique, it can be easily seen that the additional term in the
Hamiltonian $\tilde H$ enters  in the one-particle Green's
function as a spin-dependent imaginary chemical potential, which
alternatively may be absorbed in the Matsubara frequencies
$$
\omega_n=2\pi T(n+[1+\alpha\sigma]/2)
$$
with $\sigma=\pm 1$. These spin-dependent frequencies interpolate
between the bosonic and fermionic result and constitute the
semi-fermionic description of the model.

The mean-field result for the occupation probability $\langle
n_{i\sigma}\rangle$ wit the chemical potential $\mu =\mu_0+U-4 z
t^2/U \nu$ ($z$ is a coordination number, $\nu$ is a filling
factor) gives the complex value
\begin{equation}
\langle n_{i\sigma}\rangle=\frac {1}{\exp(-\beta
\mu)+1}+i\frac{\sigma}{2\cosh(\beta\mu/2)}
\end{equation}
similarly to complex distribution function derived in real-time
spin formalism (\ref{complex1}). The result reflects the fact that
he electron occupation probability $\langle n_{i\sigma}\rangle$ is
not an observable quantity in the strong coupling limit, whereas
its real part $Re \langle n_{i \uparrow}\rangle= Re \langle n_{i
\downarrow}\rangle=(\langle n_{i \uparrow}\rangle + \langle n_{i
\downarrow}\rangle)/2$ gives a bi-fermionic distribution. The
non-vanishing and purely imaginary quantity $\langle n_{i
\uparrow}\rangle -\langle n_{i \downarrow}\rangle$ describes the
magnetization of the system as a response to the fictitious
imaginary magnetic field, which is also not an observable
quantity.

 The mean-field semi-fermionic theory of
superconductivity in LNU-Hubbard model was developed in
\cite{oppermann91a,stein92}. The basic result indicates on
Bose-condensation type of the superconductivity with critical
temperature
\begin{equation}
T_c=\frac{2|t^*|}{\sqrt{z}}\left[\frac{\nu(2\nu-1)(1-\nu)^3}{\ln(\nu/(1-\nu)}\right]^{1/2}
\end{equation}
with $t^*=zt$. the critical temperature varies as $T_c\sim
(\nu\ln^{-1}\nu)^{1/2}$ for $\nu\to 0$ and $T_c\sim
|t^*|/2\sqrt{2z}$ for $\nu\to 1/2$. We address the reader to
original papers \cite{oppermann91a,stein92} for details of loop
corrections and excitation spectra in the LNU-Hubbard model.

The t-J Hamiltonian has  the form
\begin{equation}
H=-t\sum_{\langle
i,j\rangle,\sigma}(1-a^\dagger_{i,-\sigma}a_{i,-\sigma})a^\dagger_{i,\sigma}a_{j,\sigma}(1-a^\dagger_{j,-\sigma}a_{j,-\sigma})
+J\sum_{\langle i,j\rangle}\vec{S}_i\vec{S}_j \label{tJ}
\end{equation}
and can be obtained from the Hubbard model near half-filling point
under condition $t\ll U$. The spin exchange coupling $J=4t^2/U$.

The t-J model in the form (\ref{tJ}) does not posses the most
important condition for the application of semi-fermionic
representation. Namely, it satisfies neither the particle-hole
symmetry manifested for $SU(2)$ spin systems, nor global
particle-hole symmetry of LNU-Hubbard model. In the paper
\cite{gross90} Gross and Johnson proposed to consider generalized
version of t-J model when kinetic energy is chosen to be
particle-hole symmetric
\begin{equation}
T=-t\sum_{\langle
i,j\rangle,\sigma}\left[(1-a^\dagger_{i,-\sigma}a_{i,-\sigma})a^\dagger_{i,\sigma}a_{j,\sigma}(1-a^\dagger_{j,-\sigma}a_{j,-\sigma})
+
n_{i,-\sigma}a^\dagger_{i,\sigma}a_{j,\sigma}n_{j,-\sigma}\right]
\end{equation}
where $n_{i,\sigma}=a^\dagger_{i,\sigma}a_{i,\sigma}$. It has been
shown in \cite{gross90} that the thermodynamical properties of
original and the generalized t-J model are identical.

The partition function of t-J model is mapped to those of
generalized t-J model through projection operator \cite{gross90}.
Unlike particle-hole symmetrical spin case, where both empty and
double occupied states have to be excluded in calculation of
traces, only doublons (double occupied states) are unwanted in t-J
model. The exclusion of double occupied states is not done by
Schwinger-fermion procedure when the constraint
$n_{i\uparrow}+n_{i\downarrow}=1$ is enforced by an additional
$U(1)$ fluctuating field $\lambda$
$$
\int_0^{2\pi} d\lambda \langle\exp\left(-\beta H
+i\lambda_i[n_{i\uparrow}+n_{i\downarrow}-1]\right)\rangle
$$
but by discrete complex Lagrange multipliers (chemical potential).
The "complex" chemical potential $\mu_{tot}$ consists of the real
part $\alpha$ which takes care about finite hole concentration and
imaginary part which represents the constraint. The auxiliary
variable $\alpha =Re\mu_{tot}$ is related to the chemical
potential of holes $\mu_h$ through equation
\begin{equation}
2\sin(\alpha\beta)=\exp(\mu_h\beta)
\end{equation}
The half-filled case is described by the limit $\alpha \to 0$. It
corresponds to the Heisenberg model for which semi-fermions are
characterized by purely imaginary chemical potential $\mu =i\pi
T/2$. Thus, the semi-fermionic representation is generalized for
non-particle-hole symmetric cases as well. Another example without
particle-hole symmetry (dynamical symmetries) will be considered
in the Section devoted to application of semi-fermions in
mesoscopic physics.

\subsection{Kondo lattices: competition between magnetic and Kondo correlations}
The problem of competition between Ruderman-Kittel-Kasuya-Yosida (RKKY)
magnetic exchange and Kondo correlations is one of the most interesting
problem of the heavy fermion physics. The recent experiments unambiguously show,
that such a competition  is responsible for many unusual properties of
the integer valent heavy fermion compounds e.g. quantum critical
behavior, unusual antiferromagnetism  and superconductivity (see references in \cite{kis02a}).
We address the reader to the review \cite{col} for details of  complex physics of Kondo effect
in heavy fermion compounds. In this section we discuss the influence of Kondo effect on the
competition between local (magnetic, spin glass) and non-local (RVB) correlations.
The Ginzburg-Landau theory for nearly antiferromagnetic Kondo lattices has been constructed in
\cite{kis02a}
using the semi-fermion approach. We discuss the key results of this theory.

The  Hamiltonian of the Kondo lattice (KL) model is given by
\begin{equation}
H=\sum_{k\sigma}\varepsilon_k c^\dagger_{k\sigma}c_{k\sigma}+
J\sum_{j}\left({\bf S}_j{\bf s}_j+\frac{1}{4}N_jn_j
\right)
\label{2.1}
\end{equation}
Here the local electron and spin density operators for conduction electrons
 at site $j$ are defined as
\begin{equation}
n_j=\sum_{j\sigma}c^\dagger_{j\sigma}c_{j\sigma},~~~
{\bf s}_j=\sum_{\sigma}\frac{1}{2}c^\dagger_{j\sigma}
{\hat\tau}_{\sigma\sigma'}c_{j\sigma'},
\label{2.2}
\end{equation}
where ${\hat\tau}$ are the Pauli
matrices and $c_{j\sigma}=\sum_k c_{k\sigma}\exp (ikj)$.
The spin glass (SG) freezing is possible if an additional quenched randomness
of the inter-site exchange $I_{jl}$ between the localized spins arises.
This disorder is described by
\begin{equation}
H'=\sum_{jl}I_{jl}({\bf S}_j{\bf S}_l).
\label{2.3}
\end{equation}
We start with a perfect Kondo lattice. The spin correlations in KL
are characterized by two energy scales, i.e., $I\sim \
J^2/\varepsilon_F,$ and $\Delta_K\sim
\varepsilon_F\exp(-\varepsilon_F/J)$ (the inter-site indirect
exchange of the RKKY type and the Kondo binding energy,
respectively). At high enough temperature, the localized spins are
weakly coupled with the electron Fermi sea having the Fermi energy
$\varepsilon_F$, so that the magnetic response of a rare-earth
sublattice of KL is of paramagnetic Curie-Weiss type. With
decreasing temperature either a crossover to a strong-coupling
Kondo singlet regime occurs at $T \sim \Delta_K$ or the phase
transition to an AFM state occurs at $T=T_N \sim zI$ where $z$ is
a coordination number in KL. If $T_N \approx \Delta_K$ the
interference between two trends results in the decrease of both
characteristic temperatures or in suppressing one of them. The
mechanism of suppression is based on the screening effect due to
Kondo interaction. As we will show, the Kondo correlations screen
the local order parameter, but leave nonlocal correlations intact.
The mechanism of Kondo screening for single-impurity Kondo problem
is illustrated on Fig.6
\begin{figure}
\begin{center}
  \epsfysize6cm \epsfbox{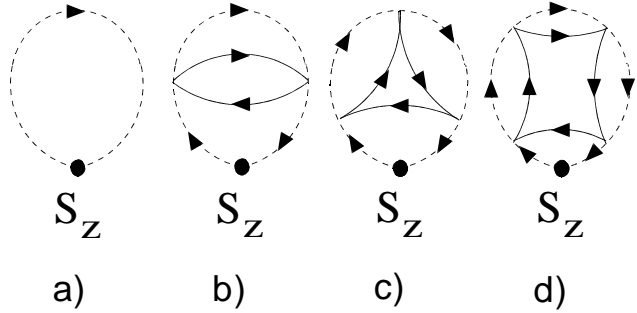}
\vspace*{-10mm}
\end{center}
\vspace*{2mm}
\caption{Kondo screening of the local moment in single-impurity Kondo problem. Dashed line denotes semi-fermions, solid line stands for conduction electrons.}
\label{fig:pol2}
\end{figure}
As a result, the magnetization of local impurity in the presence of Kondo effect is
determined in terms of GF's of semi-fermions ${\cal G}(\omega)$  by the following expression \cite{WT}:
$$
{\cal M}(H)=S(g\mu_B)T\sum_{\omega}\left({\cal G}_\uparrow(\omega)-
{\cal G}_\downarrow(\omega)\right)=S(g\mu_B)
\tanh\left(\frac{H\beta}{2}\right)\times
$$
\begin{equation}
\times \left[1-\frac{1}{\ln(T/T_K)}-\frac{\ln(\ln(T/T_K))}{2\ln^2(T/T_K)}+...\right].
\end{equation}
To take into account the screening effect in the lattice model we apply the semi-fermionic representation
of spin operators. In accordance with the general path-integral approach to KL's, we first integrate
over fast (electron) degrees of freedom. The Kondo exchange interaction is decoupled by
auxiliary field $\phi$ \cite{read83a}
with statistics complementary to that of semi-fermions which prevents this field
from Bose condensation except at $T=0$. As a result, we are left with an effective bosonic action
describing low-energy properties of KL model at high $T>T_K$ temperatures.
\begin{itemize}
\item Kondo screening of the N\'eel order
\end{itemize}
To analyze the influence of Kondo screening on formation of AFM order, we adopt the decoupling
scheme for the Heisenberg model discussed in Section II.A. Taking into account the classic part
of N\'eel field, we calculate the Kondo-contribution to the effective action which depends on magnetic
order parameter ${\cal N}$:
\begin{equation}
{\cal A}_{\phi}=2\sum_{{\bf q},n}
\left[\frac{1}{\widetilde J}-\Pi({\cal N})\right]
|\phi_n({\bf q})|^2.
\label{3.1}
\end{equation}
where a polarization operator $\Pi({\cal N})$ casts the form
\begin{equation}
\Pi({\cal N})=\rho(0)\ln\left(\frac{\epsilon_F}{T}\right)+\left[
\frac{\pi}{2}\left(\frac{1}{\cosh(\beta{\cal N})}-1\right)
+O\left(\frac{{\cal N}^2}{T\epsilon_F}\right)\right]~,
\label{3.2}
\end{equation}
where $\rho(0)$ is the density of states of conduction electrons at the Fermi level and the
Kondo temperature $T_K=\epsilon_F\exp\left(-1/(\rho(0)J)\right)$.
Minimizing the effective action ${\cal A}(\phi, {\cal N})$ with respect
to classic field ${\cal N}$, the mean field equation
for N\'eel transition is obtained (c.f. with (\ref{n1}))
\begin{equation}
{\cal N}=\tanh\left(\displaystyle\frac{I_{\bf Q}{\cal N}}{2T}\right)
\left[\displaystyle
1-\frac{a_N}{\displaystyle\ln\left(T/T_K\right)}
\frac{\cosh^2(\beta I_{\bf Q}{\cal N}/2)}
{\cosh^2(\beta I_{\bf Q} {\cal N})}\right].
\label{3.4}
\end{equation}
As a result, Kondo corrections to the molecular field equation reduce the N\'eel temperature
\begin{itemize}
\item Kondo enhancement of RVB correlations
\end{itemize}
Applying the similar procedure to nonlocal RVB correlations, we take into account the influence
of Kondo effect on RVB correlations
$$
\Pi(I_{\bf q}\Delta)=\rho(0)\ln\left(\frac{\epsilon_F}{T}\right)+
$$
\begin{equation}
+\sum_{\bf k}
\left[\frac{1}{\cosh \beta(I_{\bf k}\Delta)}-1+
I_{\bf k}\Delta\tanh(\beta I_{\bf k}\Delta)\right]
\frac{1}{\xi^2_{\bf k+q}+ (\pi/2\beta)^2}~.
\label{3.6}
\end{equation}
Here $\xi_k=\epsilon({\bf k})-\epsilon_F$.
Minimizing the effective action with respect to $\Delta$ we obtain new self-consistent equation
to determine the non-local semi-fermion correlator.
\begin{equation}
\Delta= -\sum_{\bf q}\frac{I_{\bf q}}{I_0} \left[\tanh\left(\frac{I_{\bf q}\Delta}{T}
\right)+a_{sl}\frac{I_{\bf q} \Delta}{T\ln(T/T_K)}\right].
\label{3.4a}
\end{equation}
It is seen that unlike the case of local magnetic order, the Kondo scattering favors transition into
the spin-liquid state, because the scattering means the  involvement of the itinerant
electron degrees of freedom into the spinon dynamics.

\subsection{Semi-fermionic representation for spin glass models}

In this Section we sketch the technicalities  related to the
application of Popov-Fedotov method to disordered spin systems. We
address the reader to a review \cite{binder} on theoretical
concepts and experimental facts on spin glasses.

We consider Ising/Heisenberg spin glass model
\begin{equation}
H=-\sum_{<ij>}I_{ij}\left(S^z_iS^z_j +\frac{\lambda}{2}
\left[S^+_iS^-_j+S^-_iS^+_j\right]\right)+g h\sum_i S^z_i
\end{equation}
with gaussian distributed random interaction $I_{ij}$ and
$\lambda=0$ corresponds to Sherrington-Kirpatrick (SK) model
\cite{SK}, while $\lambda=1$ corresponds to isotropic Heisenberg
model.

Following the procedure described in
\cite{oppermann92,oppermann93}  we use fermionic representation of
spin operators employing the replica trick \cite{oppermann92}:
\begin{equation}
a_i(\tau)\;\;\;\;\;\; \to \;\;\;\;\;\; \varphi^a_i(\tau),
\;\;\;\;a=1..n.
\end{equation}
 As a result we obtain the expression for the
average of $n$-th power of the partition function for isotropic
model:
$$\langle Z^n\rangle_{av}=\prod\int d I_{ij} P(I_{ij})\prod D[\varphi^a_{i,\sigma}]
\exp\left(\int_0^\beta d\tau\left[\sum_i \bar\varphi_{i,\alpha}^a
\left[\partial_\tau-i\pi\beta^{-1}/2\right]\varphi_{i,\alpha}^a+\right.\right.
$$
\begin{equation}\left.\left. +\sum_{<i,j>}I_{ij}\sum_{a=1}^n
\bar\varphi^a_{i,\alpha}\vec{\sigma}\varphi^a_{i,\alpha'}
\bar\varphi^a_{j,\gamma}\vec{\sigma'}\varphi^a_{j,\gamma'}\right]\right)
\end{equation}
Assuming $P(I_{ij})$ to be a gaussian distribution $P(I_{ij}) \sim
\exp (-I^2_{ij}N/(2I^2))$ we integrate over $I_{ij}$. As a result,
four-spin, or, eight-fermion  term appears in effective action. To
decouple the quantity
\begin{equation}
X_\mu^{ab}(\tau,\tau')=\sum_i\sum_{\alpha\alpha'\gamma\gamma'}
\bar\varphi^a_{i,\alpha}(\tau)\sigma_\mu\varphi^a_{i,\alpha'}(\tau)
\bar\varphi^b_{i,\gamma}(\tau')\sigma_\mu'\varphi^b_{i,\gamma'}(\tau')
\end{equation}
the $Q$-matrices are introduced:
$$\langle Z^n\rangle_{av}=\int D[\vec{Q}]\exp\left(
-\frac{1}{4}(\beta I)^2N {\tt Tr}[\vec{Q}^2]+\right.
$$
\begin{equation}
+\left. \sum_i\ln\left[\prod\int D[\varphi]\exp\left(
\sum_{a,\alpha}\bar\varphi_{i\alpha}^a({\cal
G}_0^{-1})_{\alpha\alpha} \varphi^a_{i\alpha}+\frac{1}{2}(\beta
I)^2 {\tt Tr}[\vec{Q}\vec{X}]\right)\right]\right)
\end{equation}
and $({\cal G}_0^{-1})_{\alpha\alpha'}=\delta_{\alpha\alpha'}
(i\omega_n-\alpha g h/2)$. For Ising model only $X_3^{ab}$ should
be considered.

The next step is to consider the second decoupling procedure for
${\tt Tr}[\vec{Q}\vec{X}]$. For this sake, following
\cite{oppermann92,oppermann93} we consider $Q$ as a constant
saddle-point matrix.
\begin{equation}
Q_{SP}^{aa}=\tilde q,\;\;\;\;\;\;\;Q_{SP}^{a\ne b}=q
\end{equation}
The replica-global term $(\sum_i S^a_\varphi)^2$ is decoupled by a
replica independent Hubbard-Stratonovich field, while the replica
local product $(S^a_\varphi)^2$ requires a replica-dependent
decoupling field \cite{oppermann93}. Although it is well known
that both SK and isotropic vector model of infinite range spin
glass are unstable with respect to Parisi's broken symmetry
permutation symmetry \cite{Parisi}, we shall consider first the
replica symmetrical solutions existing only near the phase
transition point.

The last step is to integrate over all Popov-Fedotov fermions
taking the limit $n\to 0$ and calculate free energy per site
\begin{equation}
f=T\lim_{n\to 0}\frac{1}{nN}\left(1-<Z^n>_{av}\right)
\end{equation}
and find the extremum of $f$ with respect to $q$ and $\tilde q$.

We emphasize, that in the framework of quantum theory treatment we
do not have a freedom to choose $\tilde q$ freely. In general, the
equations for $q$ and $\tilde q$ are the set of coupled equations.
\\

{\bf a) Sherrington-Kirkpatrick model.\\}

After applying the scheme described above to Ising spin-glass
model $(\lambda=0)$ and some straightforward calculations one gets
the following expressions:
\begin{equation}
\exp\left(\frac{1}{2}(\beta I)^2{\tt Tr}[QX]\right)=\int_z^G
\prod_a\int_{y_{a}}^G
\exp\left(\sum_{a,\alpha}\bar\varphi^a_{i\alpha}\sigma
H(z,y^a)\varphi^a_{i\alpha} \right)
\end{equation}
 where
$H(z,y^a)=I\sqrt{q}z+I\sqrt{\tilde q -q}y^a$ and $\displaystyle
\int_x^G...=1/\sqrt{2\pi}\int_{-\infty}^{+\infty}d x\exp(-x^2/2)
...$ Then, the integration over $\varphi^a$ becomes gaussian and
can be performed explicitly.

Performing the integration over $y^a$ the following expression for
the free energy is obtained:
\begin{equation}
\beta f=\frac{(\beta I)^2}{4}\left((1-\tilde
q)^2-(1-q)^2\right)-\int_z^G \ln[2\cosh(\beta I\sqrt{q}z)]
\end{equation}
resulting in the equations:
\begin{equation}
q=\int_z^G\tanh^2(\beta I\sqrt{q} z),\;\;\;\;\;\; \tilde q=1.
\label{sk1} \end{equation}
Thus, the SK results are exactly
reproduced by Popov-Fedotov method.
\\

{\bf b) Isotropic vector spin-glasses.\\}

Applying the same procedure to isotropic vector spin-glass model,
we obtain the  set of equations for diagonal $\tilde q$ and
non-diagonal $q$ saddle point elements of $Q$ matrix
\cite{oppermann92,oppermann93}:
\begin{equation}
q=\frac{2}{3q^{3/2}(\beta I)^3\sqrt{2\pi}}\int_0^\infty d r
\exp\left(-\frac{r^2}{2q(\beta
I)^2}\right)\left[\frac{b^2+\tanh(r)[r-b^2/r]}
{1+b^2\tanh(r)/r}\right]^2$$
\end{equation}

\begin{equation}
\tilde q=\frac{2}{3q^{3/2}(\beta I)^3\sqrt{2\pi}}\int_0^\infty d r
\exp\left(-\frac{r^2}{2q(\beta I)^2}\right)
\left[\frac{r^2+r\tanh(r)[b+2]} {1+b^2\tanh(r)/r}\right]
\end{equation}
with $b=\sqrt{\tilde q - q}\beta I$. These equations are the
generalization of equations Eq.\ref{sk1} for replica-symmetric
solution of isotropic Heisenberg spin glass model. We address the
readers to original papers \cite{oppermann92,oppermann93} for
discussion of stability (de Almeida-Thouless) line of the vector
model and broken replica solutions in the framework of
semi-fermionic approach.

\subsection{Kondo lattices with quenched disorder}
In this Section we consider the Kondo Lattice model (\ref{2.1})
with random RKKY interaction. The randomness is associated with
e.g. the presence of non-magnetic impurities in magnetic compound
resulting in appearance of random phase in the RKKY indirect
exchange. In this case the spin glass phase should be considered.
As it has been shown in \cite{kiselev00a} and \cite{kis02a}, the
influence of static disorder on Kondo effect in models with Ising
exchange on fully connected lattices (SK model) can be taken into
account by the mapping KL model with quenched disorder onto the
single impurity Kondo model in random (depending on replicas)
magnetic field. It allows for the self-consistent determination of
the Edwards-Anderson $q_{EA}$ order parameter given by the
following set of self-consistent equations
$$
\tilde q  =  1-\frac{2c}{\ln(T/T_K)} -
O\left(\frac{1}{\ln^2(T/T_K)}\right),
$$
\begin{equation}
q  =  \int_x^G\tanh^2\left(\frac{\beta I x \sqrt{q}} {1+2c(\beta
I)^2(\tilde q - q)/\ln(T/T_K)}
\right)+O\left(\frac{q}{\ln^2(T/T_K)}\right) ~.
\end{equation}
Here $q=q_{EA}$ and $\tilde q$ are nondiagonal and diagonal
elements of Parisi matrix respectively. Therefore, the
Kondo-scattering  results in the depression of the freezing
temperature due to the screening effects in the same way as the
magnetic moments and the one-site susceptibility are screened in
the single-impurity Kondo problem  when Ising and Kondo
interactions are of the same order of magnitude.
\begin{figure}
\begin{center}
\epsfxsize35mm \epsfbox{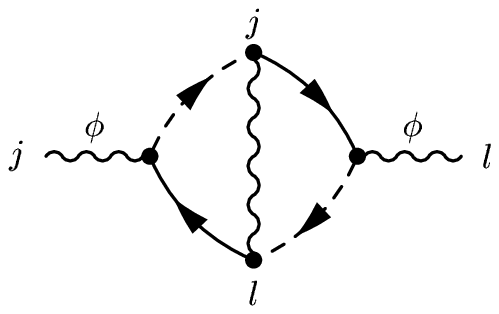} \epsfxsize35.mm
\epsfbox{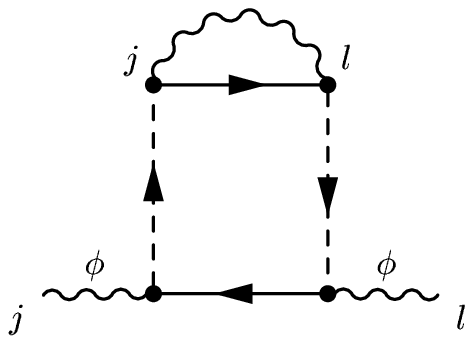} \caption{Feynman diagrams for nonlocal
excitations associated with the overlap of Kondo clouds. Wavy line
denotes Kondo "semi-Bosonic" field, solid line stands for
fermions, dashed line denotes semi-fermions.}
\end{center}
\end{figure}
Let's now briefly discuss the  fluctuation effects in Kondo
lattices. The natural way to construct the fluctuation theory is
to consider the non-local dynamical Kondo correlations described
by the field $\phi({\bf q},\omega)$ (see Fig.7). In fact, the
non-locality of the ``semi-Bosonic'' field is associated with an
overlap of Kondo clouds \cite{kis02a} and responsible for a
crossover from the localized magnetism to the itinerant-like
fluctuational spin-liquid magnetism. The temperature dependence of
static magnetic susceptibility  becomes nonuniversal in spite of
the fact that we are in a region of critical AFM fluctuations
which is consistent with recent experimental observations (see
\cite{kis02a} for details).
\section{Application of semi-fermions in mesoscopic and nano-physics}
\subsection{Kondo effect in quantum dots}
The single electron tunneling through the quantum dot \cite{Glaz}
has been  studied in great details during the recent decade. Among
many interesting phenomena behind the unusual transport properties
of mesoscopic systems, the Kondo effect in quantum dots, recently
observed experimentally, continues to attract an attention both of
experimental and theoretical communities.  The modern nanoscience
technologies allow one to produce the highly controllable systems
based on quantum dot devices and possessing many of properties of
strongly correlated electron systems. The quantum dot in a
semiconductor planar heterostructure is a confined few-electron
system (see Fig.8) contacted by sheets of two-dimensional gas
(leads). Junctions between dot and leads produce the exchange
interaction between the spins of the dot and spins of itinerant 2D
electron gas. Measuring the dc $I-V$ characteristics, one can
investigate the Kondo effect in quantum dots under various
conditions.

Various realizations of Kondo effect in quantum dots were proposed
both theoretically and experimentally in recent publications (see
e.g. \cite{KA02,KKA}  for review). In order to illustrate the
application of semi-fermionic approach we discuss briefly electric
field induced Kondo tunneling in double quantum dot (DQD). As was
noticed in \cite{KA01}, quantum dots with even ${\cal N}$ possess
the dynamical symmetry $SO(4)$ of spin rotator in the Kondo
tunneling regime, provided the low-energy part of its spectrum is
formed by a singlet-triplet (ST) pair, and all other excitations
are separated from the ST manifold by a gap noticeably exceeding
the tunneling rate $\Gamma$. A DQD with even ${\cal N}$ in a
side-bound configuration where two wells are coupled by the
tunneling $v$ and only one of them (say, $l$) is coupled to
metallic leads $(L,R)$ is a simplest system satisfying this
condition \cite{KA01,KA04}. Such system was realized
experimentally in Ref.\cite{mol95}.
\begin{figure}
\begin{center}
\epsfxsize42mm \epsfysize28mm \epsfbox{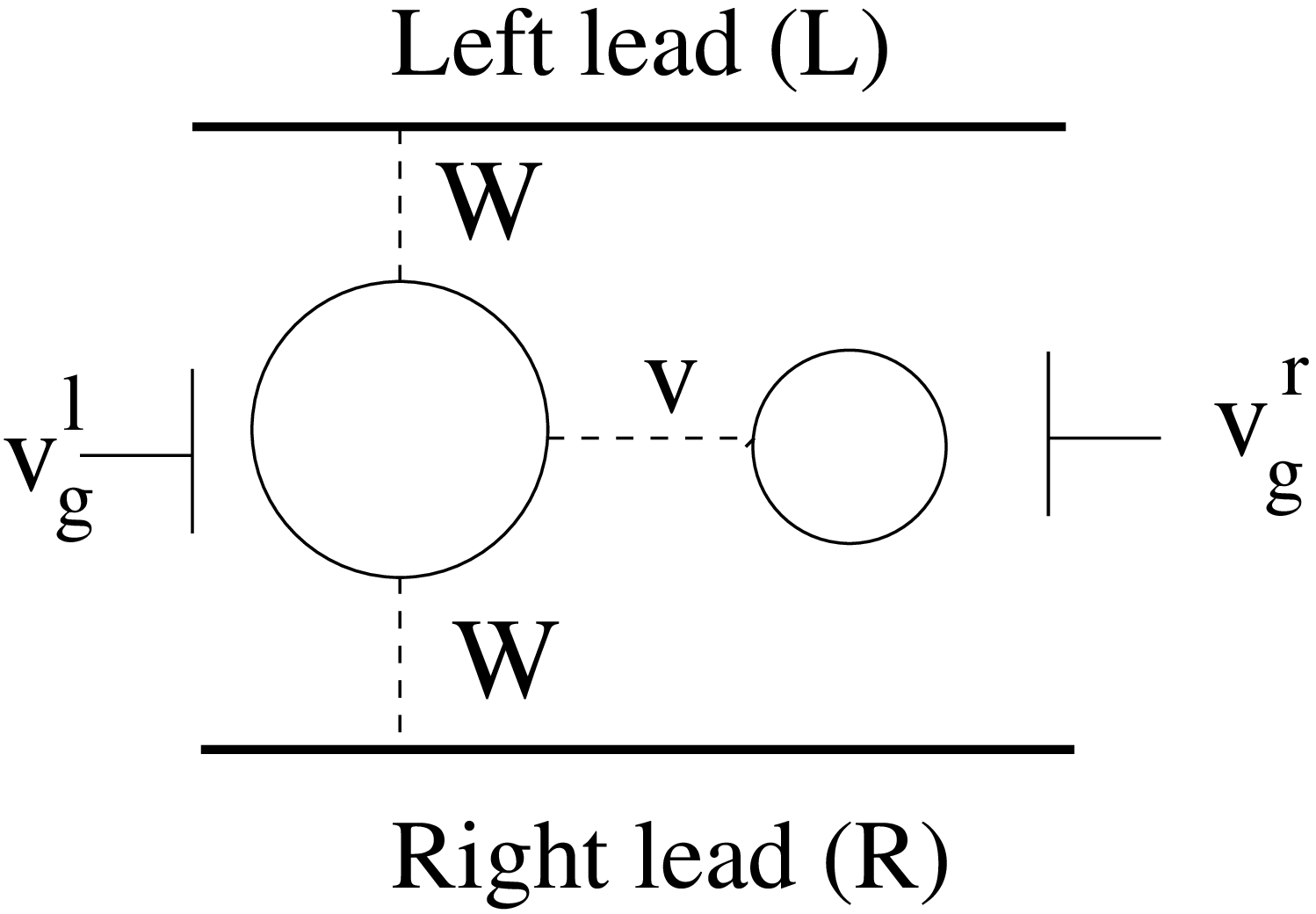} \hspace*{15mm}
\epsfxsize42mm \epsfbox{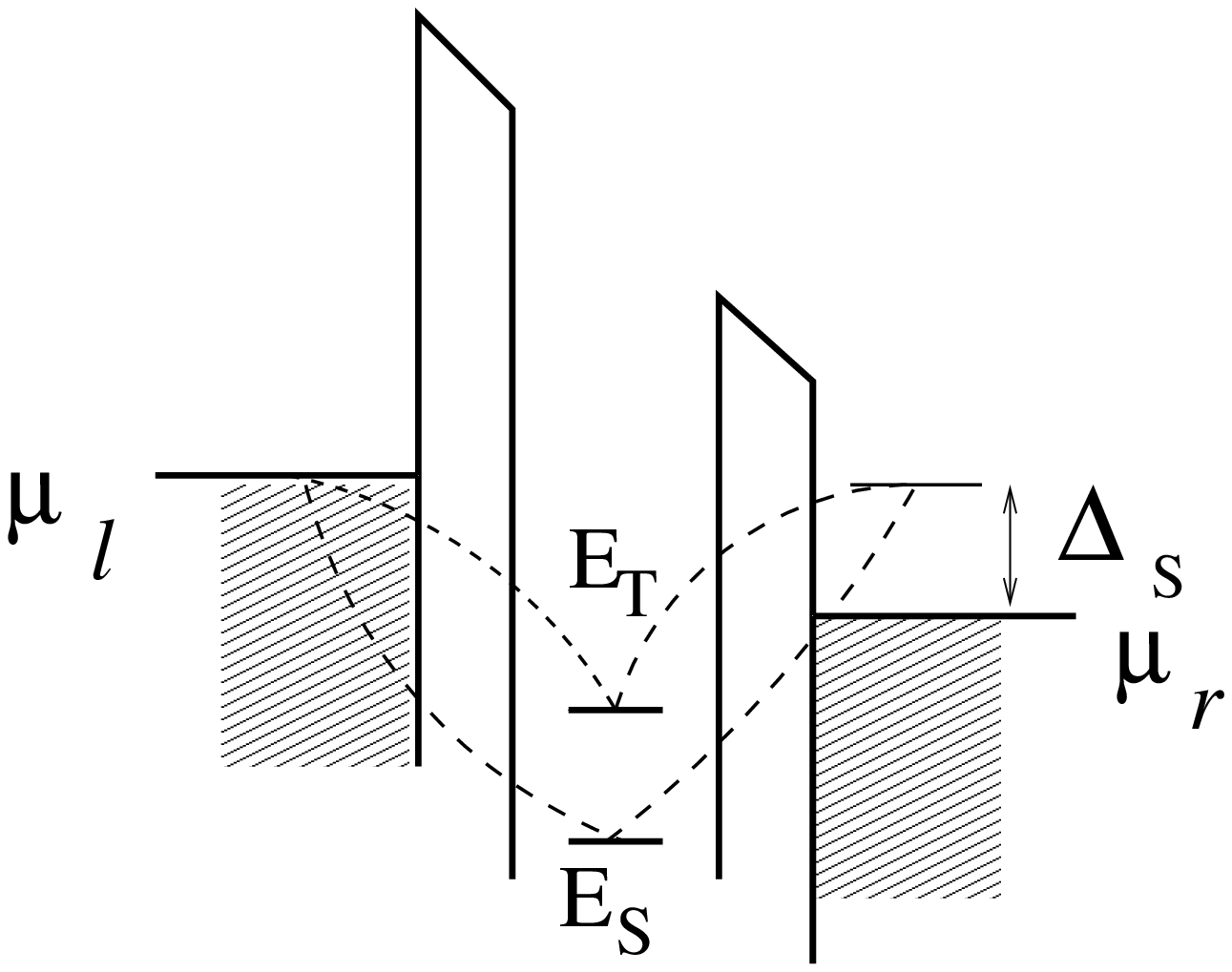}
\mbox{}\\
a)\hspace*{50mm} b)\\
\caption{(a) Double quantum dot in a side-bound configuration
(b) co-tunneling processes
in biased DQD responsible for the resonance Kondo tunneling.}
\end{center}
\end{figure}
As it was shown in \cite{kis02b} the Shrieffer-Wolff  (SW) transformation, when applied
to a spin rotator results in the following effective spin Hamiltonian
$$
H_{int}=\sum_{kk',\alpha\alpha'=L,R}J^S_{\alpha\alpha'}
f_s^\dagger f_s c^\dagger_{k\alpha\sigma}c_{k'\alpha'\sigma}
$$
\begin{equation}
+\sum_{kk',\alpha\alpha'\Lambda\Lambda'}\left(J^T_{\alpha\alpha'}
\hat S^d_{\Lambda \Lambda'} +J^{ST}_{\alpha\alpha'}\hat
R^d_{\Lambda \Lambda'}\right)\tau^d_{\sigma\sigma'}
c^\dagger_{k\alpha\sigma}c_{k'\alpha'\sigma'}f_\Lambda^\dagger
f_{\Lambda'} \label{hint}
\end{equation}
where the $c$-operators describe the electrons in the leads and
$f$-operators stand for the electrons in the dot. The matrices
$\hat S^d$ and $\hat R^d$ ($d$$=$$x$,$y$,$z$) are $4\times 4$
matrices  defined by relations (\ref{matrix}) (see Appendix A) and
$J^S=J^{SS}$, $J^T=J^{TT}$ and $J^{ST}$ are singlet, triplet and
singlet-triplet coupling SW constants, respectively.

Applying the semi-fermionic representation of $SO(4)$ group
introduced in Section I,  we started with perturbation theory
results analyzing the most divergent Feynman diagrams (Fig.9) for
spin-rotator model \cite{kis02b}.
\begin{figure}
\includegraphics[width=0.45\textwidth]{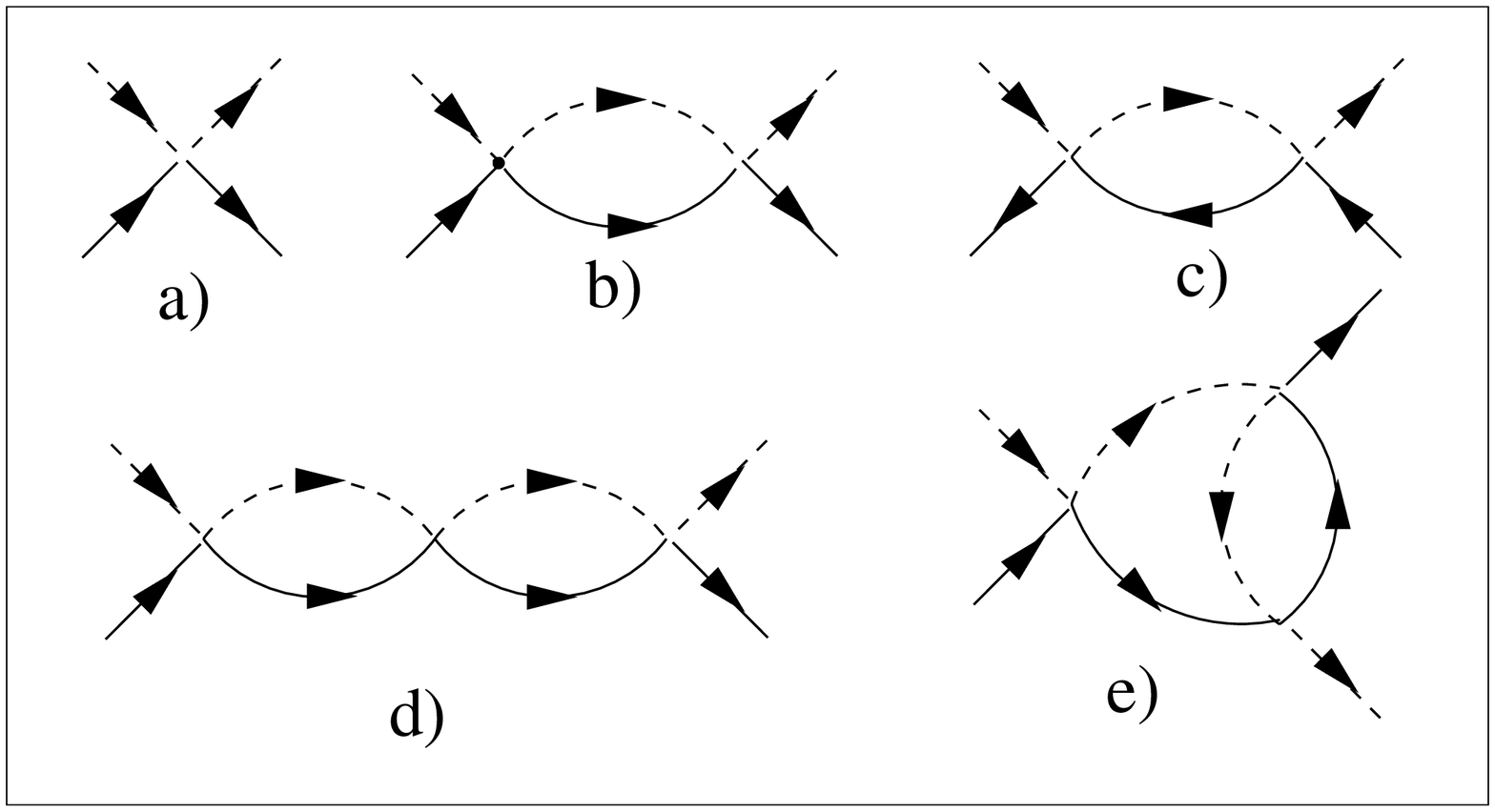}\hspace*{1cm}
\includegraphics[width=0.45\textwidth]{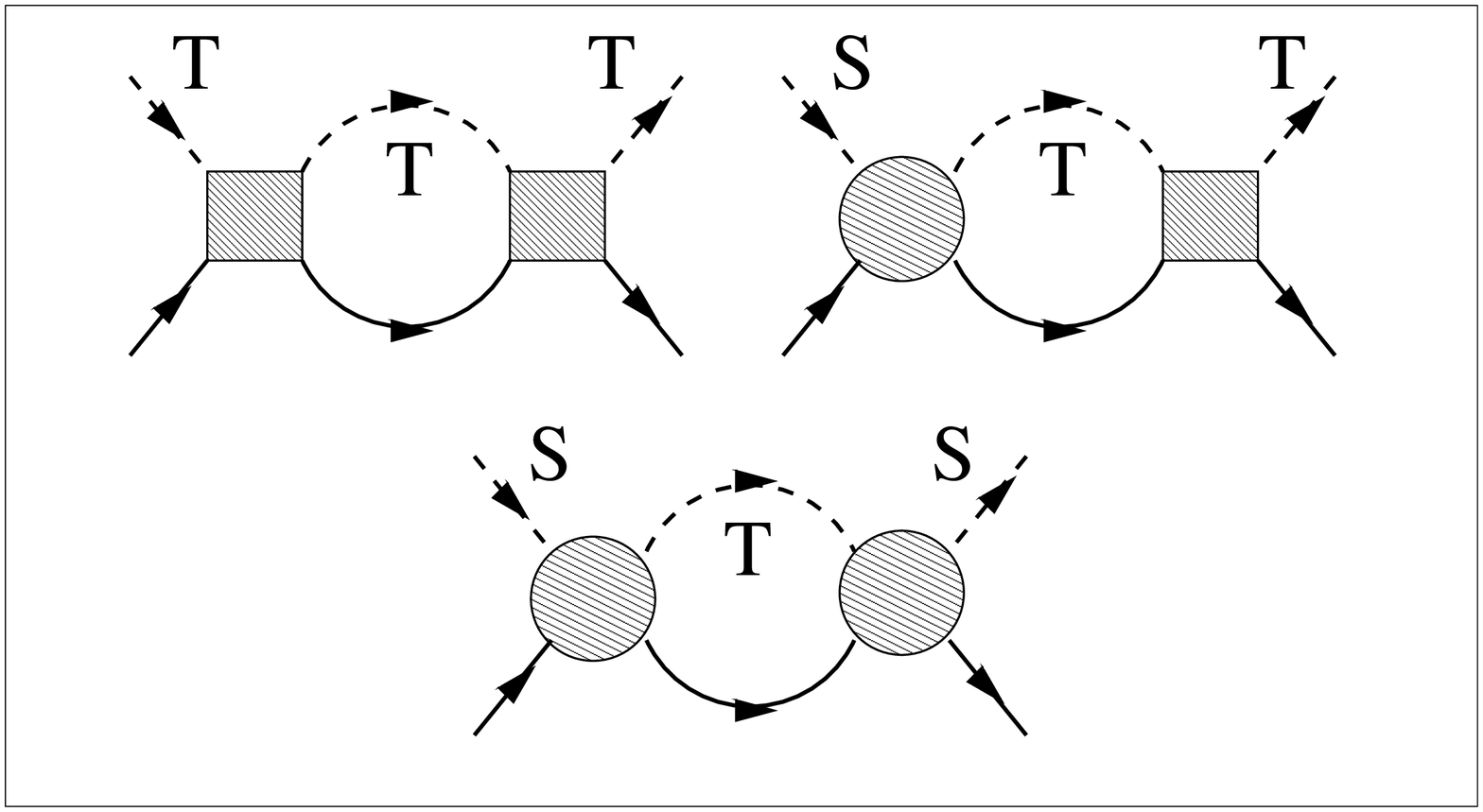}
\caption{Right panel: leading (b,d) and next to leading (c,e)
parquet diagrams determining renormalization of $J^S (a)$. Solid
lines denote electrons in the leads. Dashed lines stand for
semi-fermions representing double electron states in the dot. Left
panel: irreducible diagrams contributing to RG equations. Hatched
boxes and circles stand for triplet-triplet and singlet-triplet
vertices respectively.}
\end{figure}

Following  the "poor man's scaling" approach we derive the system
of  coupled renormalization group equations (see Fig.9) for
effective couplings responsible for the finite bias $eV$ transport
through DQD.

As a result, the differential conductance $G(eV,T)/G_0\sim
|J^{ST}_{LR}|^2$ is shown to be the universal function of two
parameters $T/T_K$ and $eV/T_K$, $G_0=e^2/\pi \hbar$:
\begin{equation}
G/G_0 \sim \ln^{-2}\left(\max[(eV-\delta),T]/T_K\right)
\label{dcond}
\end{equation}
where $T_K$ is a non-equilibrium Kondo temperature \cite{kis02b}.
Thus, the tunneling through singlet DQDs with the singlet/triplet
splitting $\delta=E_T-E_S\gg T_K$ exhibits a peak in differential
conductance at $eV\approx \delta$ instead of the usual zero bias
Kondo anomaly which arises in the opposite limit, $\delta < T_K$.
Therefore, in this case the Kondo effect in DQD is induced by a
strong external bias.

The scaling equations can also be derived in Schwinger-Keldysh
formalism (see \cite{kiselev00b} and also \cite{kis01}) by
applying the "poor man's scaling" approach directly to the dot
conductance. Thus, applying the semi-fermionic representation to
the local (zero-dimensional) problem with dynamical symmetry
\cite{KKA} we developed regular perturbation theory approach and
obtained the scaling equations.

\subsection{From double quantum dot to spin chains and ladders}
In this Section we show the application of semi-fermionic
representation to 1D systems (spin chains and ladders). The
semi-fermions give a powerful tool for fermionization of the spin
system which can be bosonized as a next step in 1D. The
bosonization procedure performed by means of Hubbard-Stratonovich
transformation \cite{yurk1,fowler,fogerby} is known in the
literature on Tomonaga-Luttinger model as "functional
bosonization" (see also \cite{yurk2,yurk3}). It is known to be
asymptotically exact in the long-wave limit. The semi-fermionic
representation allows us to apply the functional bosonization
methods to 1D spin systems \cite{kar}.

Let us consider, as an example Spin Ladder (SL) and Spin Rotator
Chain (SRC) models, Fig.10 \cite{chains}. Generic Hamiltonian for
spin systems under consideration is the Heisenberg-type spin  1/2
ladder Hamiltonian
\begin{equation}
H^{(SL)} =J_t\sum_{\langle i1,i2\rangle} {\bf s}_{i1}\cdot{\bf
s}_{i2} + J_l\sum_{\alpha}\sum_{\langle i\alpha,j\alpha\rangle}
{\bf s}_{i\alpha}\cdot{\bf s}_{j\alpha} \label{spl}
\end{equation}
Here index $\alpha=1,2$  enumerates the legs of the ladder, and
the sites $\langle i1,i2\rangle$ belong to the same rung
(Fig.10a).

A chain of dimers of localized spins illustrated by Fig. 10b is
described by the simplified version of this Hamiltonian
\begin{equation}
H^{SRC}= J_t\sum_{\langle i1,i2\rangle}{\bf s}_{i1}\cdot{\bf
s}_{i2} +J_l\sum_{\langle ij\rangle}{\bf s}_{i1}\cdot{\bf s}_{j1}
\label{1}
\end{equation}
The geometry of alternate rungs is chosen in a system (Fig.10b) to
avoid exchange interaction between spins ${\bf s}_{i2}$ and ${\bf
s}_{j2}$.

The transverse coupling may emerge either from direct exchange (in
case of localized spins) or from indirect Anderson-type exchange
induced by tunneling (similarly to the case of quantum dots). In
the latter case the sign of $J_t$ is antiferromagnetic, in the
former case it may be ferromagnetic as well. The same is valid for
$J_l$.
\begin{figure}[ht]
\includegraphics[width=0.4\textwidth]{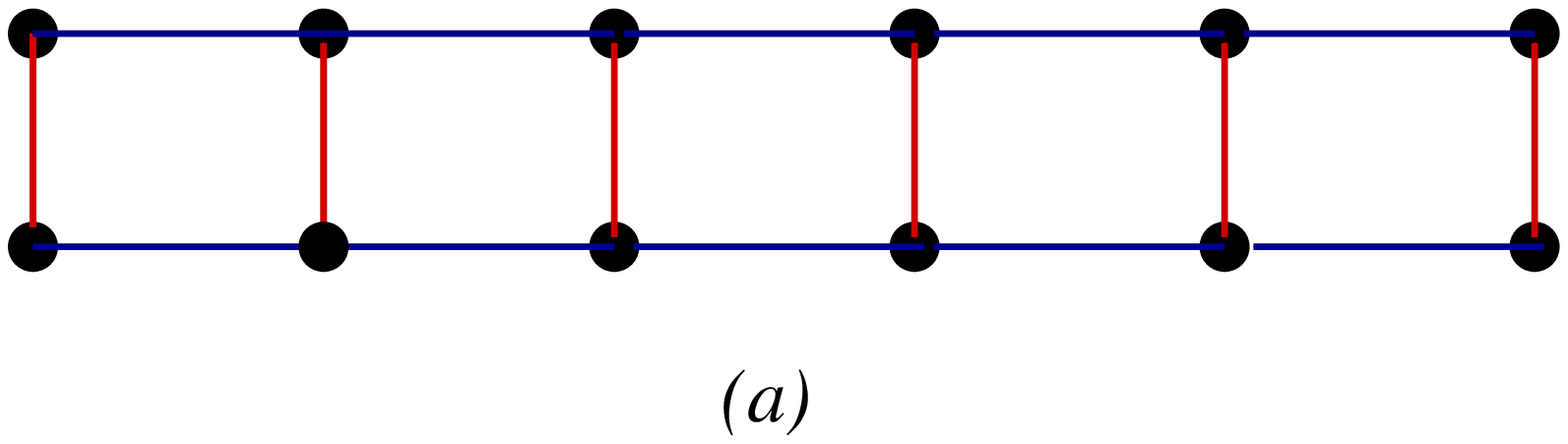}\hspace*{2cm}
\includegraphics[width=0.4\textwidth]{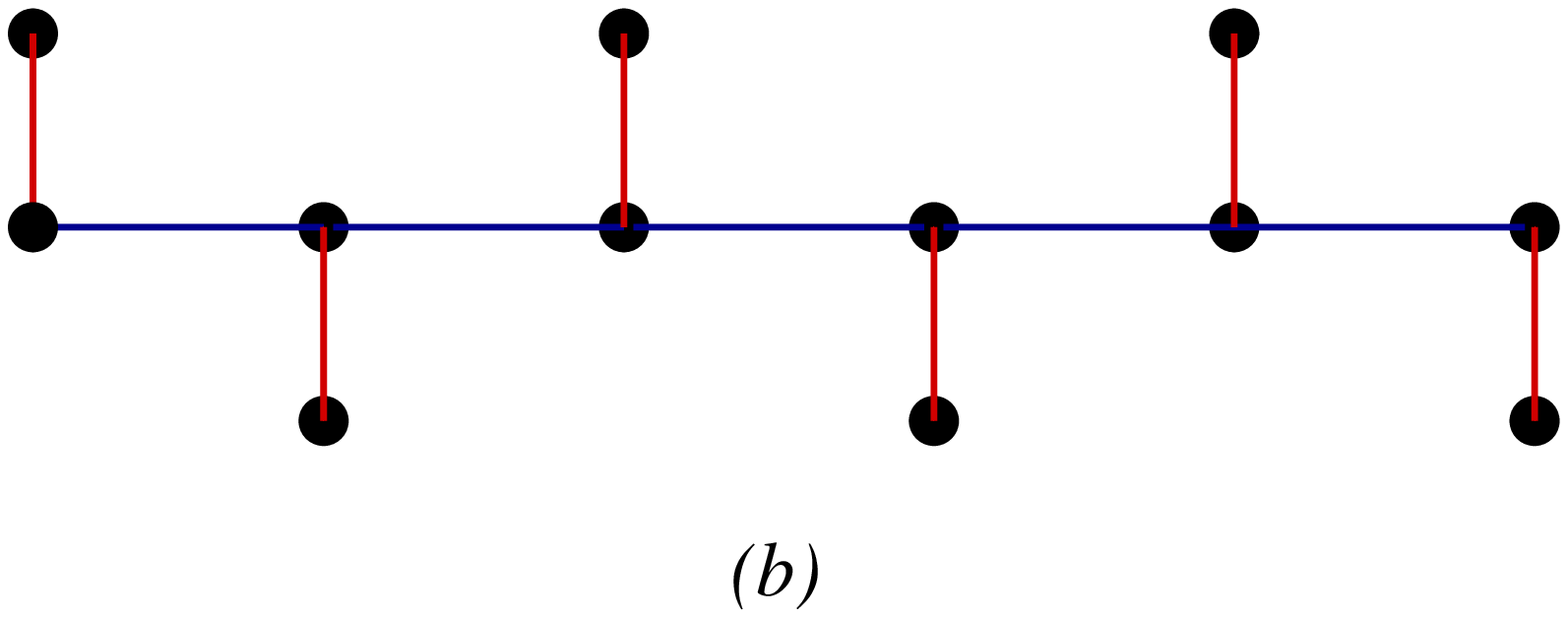}
\caption{\label{fig:nar1} Spin Ladder (a) and Spin Rotator Chain
(b)}
\end{figure}

As it is shown in \cite{kis05,chains,bort}, $SO(4)$ is an
intrinsic property of S=1/2 spin ladders and decorated spin
chains.

To describe the elementary excitations in SRC we use the
semi-fermionic representation for $SO(4)$ group (see Section I).
Then the single-site Hamiltonian may be represented in a form
\begin{equation}
H_i=-E_{ST}f_{is}^\dagger f_{is}+
h(f^\dagger_{i1}f_{i1}-f^\dagger_{i\bar{1}}f_{i\bar{1}})
\label{hi}
\end{equation}
where $h$ is uniform magnetic field applied in $z$ direction and
$E_{ST}$ is a singlet/triplet splitting.

Let us consider the XXZ-SRC model \cite{kis05}. The Hamiltonian of
this model is written in terms of the generators of $SO(4)$ group
(see Appendix B) as follows
\begin{equation}
H=\frac{1}{4}J_l\sum_{\langle
ij\rangle}\left(S^+_iS^-_j+S^+_iR^-_j+S^-_iR^+_j +\Delta[
S^z_iS^z_j+ 2S^z_iR^z_j]\right). \label{I.6}
\end{equation}

The S-S part of this Hamiltonian describe the $S=1$ chain, with
the Haldane gap in the excitation spectrum (see,
e.g.,\cite{halgap,bort}). The question is, how do the S-R
interaction modifies the gap. We consider the case of FM dimers,
when the triplet is the ground state. In this case one has one
more gap mode, where the gap equals $J_t$. This mode is coupled to
Haldane branch only via S-R exchange terms in (\ref{I.6}).

The spin liquid fermionization approach adopted here is a
convenient tool for description of Haldane spectrum. Unlike the
$S=1/2$ model, where the spin-liquid state is easily described by
globally $U(1)$ invariant modes
$T_{ij}T_{ji}=\sum_{\sigma}|f^\dagger_{i\sigma}f_{j\sigma}|^2,$ in
case of $S=1$, one deals with variables which effectively break
this symmetry. One can rewrite the effective Hamiltonian of SRC
model with $\Delta=0$ in a form
\begin{equation}
H=\frac{1}{4}J_l\sum_{ij}\left(f^\dagger_{i1}f_{j1}+
f^\dagger_{i\bar{1}}f_{j\bar{1}}\right) \bar B_j^{0S}B_i^{0S}
+f^\dagger_{i\bar{1}}f^\dagger_{j1} C_j^{0S}B_i^{0S}+ \bar
B_i^{0S}\bar C_j^{0S}f_{j1}f_{i\bar{1}}, \label{anom}
\end{equation}
where $B_j^{0S}=f_{0j}+f_{Sj}$ , $C_j^{0S}=f_{0j}-f_{Sj}$. The
terms in the first line of Eq. (\ref{anom}) describe coherent
propagation of spin fermions accompanied by a back-flow on neutral
fermions, whereas the terms in the second line are "anomalous"
(they do not conserve spin fermion number).
 For example the propagator $\langle S^+_iS^-_j\rangle$
contains anomalous components $
f^\dagger_{i1}f^\dagger_{j\bar{1}}f_{j0}f_{i0} \to
F^*_{ij,1\bar{1}}F_{ji,00}$  along with normal ones
$f^\dagger_{i1}f_{j1}f^\dagger_{j0}f_{i0}.$ Here
$F_{ij,\Lambda\Lambda'}=f_{j\Lambda}f_{i\Lambda'}$. The first term
in (\ref{anom}) describes the kinetic energy spinon excitations,
and two last anomalous term breaking U(1) symmetry are responsible
for the Haldane gap.
\begin{figure}[ht]
\includegraphics[width=0.9\textwidth]{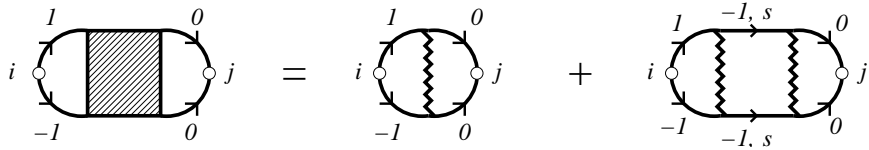}
\caption{\label{fig:anom} Lowest order contributions to anomalous
propagator.}
\end{figure}

To reveal the contribution of dynamical symmetry on the Haldane
gap, one have to note that these terms enter both in counterflow
in the first term and in gauge symmetry breaking terms in the
second line. In spin 1 ladder the counterflow term $\sim
f^\dagger_{i0}f_{j0}$ predetermines the width of spinon band
described by the first line of Eq. (\ref{anom}). Apparently, the
one more channel (tripet/singlet transitions in $B^{0S}$) enhances
this effect, because in this case the local constraint imposes
more restrictions of phase fluctuations.

The gap itself is due to anomalous correlations described by the
second line of  Eq. (\ref{anom}). Here the appearance of second
channel of spinless excitations results in formation of even and
odd operators $B_j^{0S}$ and $C_j^{0S}$. The Haldane gap closes
when the $|0 \rangle$ and $|S \rangle$ states are degenerate (the
odd operator $C_j^{0S}$ nullifies the anomalous terms responsible
for its formation). This means that appearance of $0S$ channel
{\it favors} closing of the Haldane gap.

In a strong coupling  case of $J_t\gg J_l$ both above trends may
be considered at least in the lowest order of a perturbation
theory. In case of spin ladders \cite{Dag,Barn} the 1-st and
2-nd-order  in $g=J_l/J_t$ anomalous diagrams are represented in
Fig.\ref{fig:anom}.

\section{Epilogue and perspectives}
In this paper, we demonstrated several examples of the
applications of semi-fermionic representation to various problems
of condensed matter physics. The list of these applications is not
exhaustive. We did not discuss, e.g., the interesting development
of SF approach for 2D Ising model in transverse magnetic field,
functional bosonization approach based on semi-fermionic
representation applied to $S=1/2$ chains, application of SF
formalism to mesoscopic physics \cite{col02} etc. Nevertheless, we
would like to point out some problems of strongly correlated
physics where the application of SF representation might be a
promising alternative to existing field-theoretical
methods.\\
\mbox{}\\
{\it Heavy Fermions}
\begin{itemize}
\item Crossover from localized to itinerant magnetism in Kondo lattices
\item Quantum critical phenomena associated with competition between local and nonlocal correlations
\item Nonequilibrium spin liquids
\item Effects of spin impurities and defects in spin liquids
\item Crystalline Electric Field excitations in spin liquids
\item Dynamic theory of screening effects in Kondo spin glasses.
\end{itemize}
{\it Mesoscopic systems}
\begin{itemize}
\item Nonequilibrium Kondo effect in Quantum Dots
\item Two-channel Kondo in complex multiple dots
\item Spin chains, rings and ladders
\item Nonequilibrium spin transport in wires
\end{itemize}
Summarizing, we constructed a general concept of semi-fermionic
representation for $SU(N)$ and $SO(N)$ groups. The main advantage
of this representation in application to the strongly correlated
systems in comparison with another methods is that the local
constraint is taken into account exactly and the usual Feynman
diagrammatic codex is applicable. The method proposed allows us to
treat spins on the same footing as Fermi and Bose systems. The
semi-fermionic approach can be helpful for the description of the
quantum systems in the vicinity of a quantum phase transition
point and for the nonequilibrium spin systems.
\section*{Acknowledgments}
I am grateful to my colleagues D.N.Aristov, F.Bouis, H.Feldmann,
K.Kikoin and R.Oppermann for fruitful collaboration on different
stages of SF project. This work was supported by a Heisenberg
Fellowship of Deutsche Forschungsgemeinschaft and SFB-410 project.
My special thank to participants of Strongly Correlated Workshops
in Trieste and especially to A.Protogenov for many inspiring
discussions.

\appendix{SO(N) group}
Lie algebras $o(n)$ are defined  on a basis of
$\frac{1}{2}n(n-1)$ operators of infinitesimal rotation
\begin{equation}
D_{\alpha \beta}=-D_{\beta\alpha}=x_\beta \frac{\partial}{\partial
x_\alpha}-x_\alpha \frac{\partial}{\partial x_\beta},
\;\;\;\;\;(\alpha<\beta=1,2,...n)\label{DD}
\end{equation}
which possess the following commutation relations
\begin{equation}
[D_{\alpha\beta} D_{\mu\nu}]=(\delta_{\alpha\mu}
D_{\beta\nu}-\delta_{\alpha\nu} D_{\beta\mu}-\delta_{\beta\mu}
D_{\alpha\nu}+\delta_{\beta\nu} D_{\alpha\mu}).\label{cc1}
\end{equation}
(see, e.g., \cite{Eng,Wyb}).

We consider $o(4)$ algebra as a first example of dynamical
symmetries \cite{Eng}. The antisymmetric tensor $D_{\alpha\beta}$
for $o(4)$ algebra acts in 4-dimensional space. It can be
parameterized in terms of two vectors $\bf {L}$ and $\bf {M}$ as
follows
\begin{eqnarray}
-i \left(
\begin{array}{cccc}
  0& L_3& -L_2 & M_1 \\
  &  0& L_1& M_2 \\
  & & 0& M_3 \\
  & & & 0
\end{array}
\right)
\end{eqnarray}
where the infinitesimal operators of $SO(4)$ group \cite{Wyb} in
4-dimensional space $(x,y,z,t)$ are given by
\begin{eqnarray}
L_1&=&i \left(y\frac{\partial}{\partial
z}-z\frac{\partial}{\partial y}\right),\;\; L_2=i\left(
z\frac{\partial}{\partial x}-x\frac{\partial}{\partial
z}\right),\;\; L_3=i \left(x\frac{\partial}{\partial
y}-y\frac{\partial}{\partial
x}\right),\;\;\nonumber\\
M_1&=& i\left(t\frac{\partial}{\partial
x}-x\frac{\partial}{\partial t}\right),\;\; M_2= i\left(
t\frac{\partial}{\partial y}-y\frac{\partial}{\partial
t}\right),\;\; M_3=i \left(t\frac{\partial}{\partial
z}-z\frac{\partial}{\partial t}\right).
\end{eqnarray}
From here we deduce the commutation relations:
\begin{equation}
[L_j,L_k]  =
ie_{jkl}L_l,\;\;\;\;\;\;[M_j,M_k]=ie_{jkl}L_l,\;\;\;\;\;\;\;\;\;[M_j,L_k]=ie_{jkl}M_l.
\label{com1}
\end{equation}
The two Casimir operators consist of nontrivial
\begin{equation}\label{casim}
\bf{L}^2+\bf{M}^2=3
\end{equation}
and trivial  orthogonality condition
\begin{equation}\label{ortho}
\bf{L}\cdot\bf{M}=\bf{M}\cdot\bf{L}=0.
\end{equation}

It is known, that three generators of $SU(2)$ group together with
the Casimir operator $\bf{L}^2$ define a sphere $S_2$ where each
state is parameterized by two angles. The coherent states of
$SU(2)$ group may be constructed \cite{Per} by making a standard
stereographical projection of the sphere from its south pole to
the complex plane $z$. The space of generators of $SO(4)$ group is
6-dimensional, while 2 constraints determine 4-dimensional
surface, where each state is characterized by four angles. The
stereographical  projection of this surface on a 4-D complex
hyperplane  allows to construct coherent states for $SO(4)$ group.

 The commutation relations
(\ref{com1}) can be transformed into another form by making the
linear transformation to the basis
\begin{equation}\label{comb}
J_i=\frac{L_i+M_i}{2},\;\;\;\; K_i=\frac{L_i-M_i}{2}
\end{equation}
giving more simple commutation relations
\begin{equation}
[J_j,J_k]  =
ie_{jkl}J_l,\;\;\;\;\;\;[K_j,K_k]=ie_{jkl}K_l,\;\;\;\;\;\;\;\;\;[K_j,J_k]=0.
\label{com2}
\end{equation}
The operators $L_i, M_i$ as well as $J_i, K_i$  form the elements
of the Lie algebra $o(4)$. The operators $(J_1,J_2,J_3)$ and
$(K_1,K_2,K_3)$ are separately closed under commutations, each
describing a subalgebra of $o(4)$, namely $o(3)=u(2)$. The Lie
algebra $o(4)$ is the direct sum of two $o(3)$ algebras. This
splitting of the $o(4)$ algebra into two $o(3)$ subalgebras is
directly associated with the local isomorphism between the Lie
group $SO(4)$ with the direct product group $SU(2)\times SU(2)$.
The triads $(J_1,J_2,J_3)$ and $(K_1,K_2,K_3)$ each form proper
ideals \cite{Wyb} in $o(4)$, and the Lie algebra $o(4)$ is
semi-simple.

The symmetry group of {\it spin rotator} is defined in close
analogy with the above construction, but all rotations are
performed in a spin space. The triplet/singlet pair is formed in a
simplest case by two electrons represented by their spins $s=1/2$.
 Let us  denote them as $\vec{s}_1$ and $\vec{s}_2$. The components
of these vectors obey the commutation relations
\begin{equation}
[s_{1j},s_{1k}]  =
ie_{jkl}s_{1l},\;\;\;\;\;\;[s_{2j},s_{2k}]=ie_{jkl}s_{2l},\;\;\;\;\;\;\;\;\;[s_{1j},s_{2k}]=0
\label{com3}
\end{equation}
In similarity with (\ref{com2}) these vectors may be qualified as
generators of $o(4)$ algebra, which represents a  spin rotator.
Then, the linear combinations
\begin{equation}
S_i=s_{1i}+s_{2i},\;\;\;\; R_i=s_{1i}-s_{2i}
\end{equation}
are introduced in analogy with (\ref{comb}), which define 6
generators of SO(4) group possessing the commutation relations
(\ref{com1}). These generators are represented in terms of the
Pauli-like matrices as follows
\begin{eqnarray}
S^+=\sqrt{2} \left(
\begin{array}{cccc}
  0& 1& 0 & 0\\
  0& 0 & 1& 0 \\
  0& 0& 0& 0 \\
  0& 0& 0& 0
\end{array}
\right),\;\;\;\; S^-=\sqrt{2} \left(
\begin{array}{cccc}
  0& 0& 0 & 0\\
  1& 0 & 0& 0 \\
  0& 1& 0& 0 \\
  0& 0& 0& 0
\end{array}
\right), \nonumber
\end{eqnarray}
\begin{eqnarray}
 S^z=\left(
\begin{array}{cccc}
  1& 0& 0 & 0\\
  0& 0 & 0& 0 \\
  0& 0& -1& 0 \\
  0& 0& 0& 0
\end{array}
\right),\;\;\;\; R^z=-\left(
\begin{array}{cccc}
  0& 0& 0 & 0\\
  0& 0 & 0& 1 \\
  0& 0& 0& 0 \\
  0& 1& 0& 0
\end{array}
\right),
\end{eqnarray}
\begin{eqnarray}
R^+=\sqrt{2} \left(
\begin{array}{cccc}
  0& 0& 0 & 1\\
  0& 0 & 0& 0 \\
  0& 0& 0& 0 \\
  0& 0& -1& 0
\end{array}
\right),\;\;\;\; R^-=\sqrt{2} \left(
\begin{array}{cccc}
  0& 0& 0 & 0\\
  0& 0 & 0& 0 \\
  0& 0& 0& -1 \\
  1& 0& 0& 0
\end{array}
\right). \label{matrix}
\end{eqnarray}
where the ladder operators $S^\pm = S^x \pm iS^y$, $R^\pm =R^x \pm
iR^y$. The constraints ({\ref{casim}}), (\ref{ortho}) now acquire
the form
$$
{\bf S\cdot R}=0,\;\;\;\;\;\; {\bf S}^2+ {\bf R}^2=3.
$$
 By construction $\bf{S}$
is the operator of the total spin of pair, which can take values
$S=0$ for singlet and $S=1$ for triplet states. The second
operator $\bf{R}$ is responsible for transition between singlet
and triplet states. Thus we come to the dynamical group $SO(4)$
for spin rotator.

Similar procedure is used for the $SO(5)$ group. The corresponding
$o(5)$ algebra has 10 generators
$D_{\alpha\beta}=-D_{\beta\alpha}$ (\ref{DD}) satisfying
commutation relations (\ref{cc1}). These 10 generators may be
identified as 3 vectors and a scalar in a following fashion
\begin{eqnarray}
-i \left(
\begin{array}{ccccc}
  0& S^z& -S^y & R^x & P^x \\
  &  0& S^x& R^y & P^y \\
  & & 0& R^z & P^z \\
  & & & 0 & A\\
  & & & &0
\end{array}
\right)\label{s051}
\end{eqnarray}
where the operators $\bf{S},\bf{R},\bf{P}$ and the scalar operator
$A$ obey the following commutation relations
\begin{eqnarray}
\left[S_j,S_k\right]&=&
ie_{jkl}S_l,\;\;\;\;\;\;[R_j,R_k]=ie_{jkl}S_l,\;\;\;\;\;\;\;\;\;[P_j,P_k]=ie_{jkl}S_l,\nonumber\\
\left[R_j,S_k\right]&=&
ie_{jkl}R_l,\;\;\;\;\;\;[P_j,S_k]=ie_{jkl}P_l,\;\;\;\;\;\;\;\;\;[R_j,P_k]=i\delta_{jk}T,\nonumber\\
\left[P_j,A\right]&=&
iR_l,\;\;\;\;\;\;\;\;\;\;\;\;[A,R_j]=iP_j,\;\;\;\;\;\;\;\;\;\;\;\;\;\;\;\;\;
[A,S_j]=0. \label{com5}
\end{eqnarray}
The  operators $\bf{R}$ and $\bf{P}$ are orthogonal to $\bf{S}$,
while the Casimir operator is ${\cal K}={\bf S}^2+{\bf R}^2+{\bf
P}^2+A^2=4$. These operators act in 10-D spin space, and the
kinematical restrictions reduce this dimension to 7. Similarly to
$SO(4)$ group, the vector operators describe spin S=1 and
transitions between spin triplet and two singlet components of the
multiplet, whereas the scalar $A$ stands for transitions between
two singlet states.

The group $SO(5)$ is non-compact, and the parametrization
(\ref{s051}) is not unique. As an alternative, one may refer to
another representation of $D_{\alpha\beta}$ used, the theory of
high-T$_c$ superconductivity \cite{zhang}:
\begin{eqnarray} -i\left(
\begin{array}{ccccc}
  0& &  & &  \\
  \pi_x+\pi_x^\dagger  & 0&  &  \\
  \pi_y+\pi_y^\dagger& -S^z& 0&  &  \\
  \pi_z+\pi_z^\dagger&S^y & -S^x& 0 & \\
  Q& -i(\pi_x^\dagger-\pi_x)& -i(\pi_y^\dagger-\pi_y)& -i(\pi_z^\dagger-\pi_z)&0
\end{array}
\right)\label{s052}
\end{eqnarray}
with 10 generators identified as a scalar $Q$ and three vectors
$\vec{S},\vec{\pi}$ and $\vec{\pi}^\dagger$ standing for the total
charge, spin and $\pi$ triplet $S=1$ superconductor order
parameter. Both representations (\ref{s051}) and (\ref{s052}) are
connected by the unitary transformation.

\appendix{Dynamical Symmetries in Spin Rotator Chain Model}
The Spin Ladder and Spin Rotator Chain Hamiltonians are given by
\begin{equation}
H^{(SL)} =J_t\sum_{\langle i1,i2\rangle} {\bf s}_{i1}\cdot{\bf
s}_{i2} + J_l\sum_{\alpha}\sum_{\langle i\alpha,j\alpha\rangle}
{\bf s}_{i\alpha}\cdot{\bf s}_{j\alpha}
\end{equation}

\begin{equation}
H^{SRC}= J_t\sum_{\langle i1,i2\rangle}{\bf s}_{i1}\cdot{\bf
s}_{i2} +J_l\sum_{\langle ij\rangle}{\bf s}_{i1}\cdot{\bf s}_{j1}
\label{1}
\end{equation}

We start with diagonalization of the Hamiltonian of
perpendicularly aligned dimer. The $SO(4)$ symmetry stems from the
obvious fact that the  spin spectrum of a dimer $\{i1, i2\}$ is
formed by the same singlet-triplet (ST) pair as the spin spectrum
of DQD studied in the previous section. This analogy prompts us a
canonical transformation connecting two pairs of spin vectors,
$\{{\bf s}_{i1},{\bf s}_{i2}\}$ and $\{{\bf S}_{i},{\bf R}_{i}\}$:
Two sets of spin operators are connected by a canonical
transformation
\begin{equation}
{\bf s}_{i1}=\frac{{\bf S}_{i}+{\bf R}_{i}}{2},~~~ {\bf
s}_{i2}=\frac{{\bf S}_{i}-{\bf R}_{i}}{2}, \label{rot}
\end{equation}
Then the Hamiltonian ${\cal H}_i$ of a single dimer $i$ is the
same as the Hamiltonian (\ref{hint}) of DQD. The total spin of a
dimer is not conserved in a spin chain, so the dynamical symmetry
of individual rung is revealed by the modes propagating along the
chain \cite{Barn}. Applying the rotation operation  to the
Hamiltonians (\ref{spl}), (\ref{1}), we transform them to a form
\begin{equation}
{\cal H}={\cal H}_0 +{\cal H}_{int} \label{1.1}
\end{equation}
Here
$${\cal H}_{0}=\sum_{i}{\cal H}_{i},
$$
is common for both models. It is useful to include  the Zeeman
term in ${\cal H}_i$,
\begin{equation}
{\cal H}_i=\frac{1}{2}\left(E_S {\bf R}_i^2 + E_T {\bf
S}_i^2\right) + hS_{iz}. \label{1.3}
\end{equation}
We confine ourselves by a charge sector $N_i=2$ and omit the
Coulomb blockade term for the sake of brevity. The interaction
part of SL Hamiltonian transforms under rotation (\ref{rot}) to
the following expression
\begin{equation}
{\cal H}_{int}^{SL}= \frac{1}{4}J_l\sum_{\langle ij\rangle}({\bf
S}_i{\bf S}_j+{\bf R}_i{\bf R}_j) \label{I.2a}
\end{equation}
The interaction part of the SRC Hamiltonian is
\begin{equation}
{\cal H}_{int}^{SRC}= \frac{1}{4}J_l\sum_{\langle ij\rangle}({\bf
S}_i{\bf S}_j+2{\bf R}_i{\bf S}_j+{\bf R}_i{\bf R}_j) \label{I.2}
\end{equation}

Now we see that both effective Hamiltonians belong to the same
family. In all three case one may transform initial ladder or
"semi-ladder" Hamiltonians into really one-dimensional spin-chain
Hamiltonians, which, however, takes into account the hidden
symmetry of a dimer. The effective Hamiltonians (\ref{I.2a}),
(\ref{I.2})) contain operators ${\bf R}$ describing the dynamical
symmetry of dimers. It is clear that this dynamical symmetry make
the spectrum of this Hamiltonians more rich than that of a
standard Heisenberg chain. Like in many other cases, rotation
transformation eliminates the antisymmetric combination of two
generators.

The transformation (\ref{rot}) reveals the hidden symmetry of well
known spin 1/2 ladder (\ref{I.2a}). It maps the ladder hamiltonian
onto a pair of coupled chain Hamiltonians: one is conventional
spin 1 Haldane chain , another is unconventional pseudospin 1
Haldane chain. Spin and pseudospin are coupled dynamically by the
commutation relations and kinematically by the local Casimir
constraint
\begin{equation}
{\bf S}_i^2+ {\bf R}_i^2 =3. \label{cas}
\end{equation}
One may also compare the Hamiltonian (\ref{I.2a}), (\ref{I.2})
with effective Hamiltonian of spin-1 chain, which arises after
decomposition of spin-one operators into pair of spin 1/2
operators, ${\bf S}_i= {\bf s}_i + {\bf r}_i$. This decomposition
operation transforms initial Hamiltonian into a form similar to
$H^{SRC}$ but for spin-one-half operators ${\bf s}_i,~{\bf r}_i$.
The difference between two cases is that these effective spins
commute unlike operators ${\bf S}_i,~{\bf R}_i$. In other words,
the difference is that the local symmetry of spin-one chain is
$SO(3)$ whereas the local symmetry of SRC is $SO(4)$. The spin
rotator chain (\ref{I.2}) is in some sense intermediate between
spin chains and spin ladders. In this cases the spin-pseudospin
symmetry is obviously broken by the cross terms $2{\bf S}_i{\bf
R}_j$, and our aim is to find the specific features of this novel
model systems.

In all cases the simplified versions of Heisenberg Hamiltonians
may be considered. The simplified SL  models are well known
\cite{Dag}. The simplified anisotropic versions of the Hamiltonian
(\ref{I.2}) have the following forms:

{\it Ising-like SRC model:}
\begin{equation}
H=\frac{1}{4}J_l\sum_{\langle
ij\rangle}(S^z_iS^z_j+2S^z_iR^z_j+R^z_iR^z_j) \label{I.1}
\end{equation}

{\it Anisotropic SRC model:}
$$
H=\frac{1}{4}J_l\sum_{\langle
ij\rangle}\left[(S^+_iS^-_j+S^+_iR^-_j+S^-_iR^+_j+R^+_iR^-_j)+\Delta(S^z_iS^z_j+2S^z_iR^z_j+R^z_iR^z_j)\right]
$$
{\it SRC in strong magnetic field:} SO(4) group is reduced to
SU(2) group in strong magnetic field, when the Zeeman splitting
{\it exactly} compensates the exchange gap in a single dimer,
$h_0=|E_T-E_S|$. Then at low T, the states $|i0\rangle$ and
$|i-1\rangle$ are quenched, and only two components, $R^\pm$
survive in the whole manifold . As a result, the Hamiltonian
(\ref{I.2}) is mapped onto a $XY$-model for spin one half:
\begin{equation}
H_{XY}^{(P)}= \frac{1}{4}J_l\sum_{\langle ij\rangle}(R_i^+R_j^- +
H.c.). \label{I.4}
\end{equation}
This means that starting from a singlet ground state for
$J_t\equiv E_T-E_S>0$, one may induce development of spin
liquid-like excitations by applying strong magnetic field. In a
near vicinity of this point of degeneracy, $H_{int}$ acquire the
feature of {\it XY model in transverse magnetic field}.

\end{document}